\documentclass[12pt]{article}
\pdfoutput=1
\usepackage{graphicx}
\usepackage{bm}
\def\be{\begin{equation}}
\def\ee{\end{equation}}
\def\bea{\begin{eqnarray}}
\def\eea{\end{eqnarray}}
\usepackage{latexsym}
\usepackage{epsfig,amssymb,euscript}
\usepackage{amsmath,amsfonts,bbm,mathtools}
\usepackage[T1]{fontenc}
\usepackage[utf8]{inputenc}
\usepackage{cite}
\usepackage{enumitem}
\usepackage[margin=3cm]{geometry}
\usepackage{varwidth}
\usepackage{subdepth}
\usepackage{color}
\usepackage{hyperref}
\usepackage{graphicx}
\usepackage[small,bf,font=small,textfont=it,labelsep=period]{caption}
\numberwithin{equation}{section}

\newcommand{\nn}{\nonumber\\}

\DeclareRobustCommand{\bfrac}[2]{%
	\mathchoice{\frac{\raisebox{-0.4ex}{$#1$}}{\raisebox{0.ex}{$#2$}}}%
	{\frac{\raisebox{-0.1ex}{$\scriptstyle#1$}}{\raisebox{0.ex}{$\scriptstyle#2$}}}%
	{\frac{#1}{#2}}%
	{\frac{#1}{#2}}%
}

\newcommand{\alphap}{\ensuremath{\alpha\raisebox{-0.3ex}{$'$}}}

\usepackage{array,calc,epsfig}

\usepackage{color}

\newcommand{\Tr}{{\rm Tr}}

\renewcommand{\l}{\lambda}

\usepackage{graphicx}

\def\be{\begin{equation}}
\def\ee{\end{equation}}
\def\ba{\begin{eqnarray}}
\def\ea{\end{eqnarray}}
\def\nb{\nonumber}
\def\p{\partial}

\def\a{\alpha}
\def\b{\beta}

\def\f{\varphi}
\def\ff{\phi}

\def\g{\gamma}

\def\d{\delta}
\def\D{\Delta}
\def\l{\lambda}

\def\m{\mu}
\def\n{\nu}

\def\s{\sigma}

\def\t{\tau}

\def\tt{\tilde}

\def\q{\quad}

\def\etaprime{\eta '}

\newcommand{\pr}[1]{\left(#1\right)}
\newcommand{\pq}[1]{\left[#1\right]}

\begin{document}
\baselineskip=15.5pt
\pagestyle{plain}
\setcounter{page}{1}
\newfont{\namefont}{cmr10}
\newfont{\addfont}{cmti7 scaled 1440}
\newfont{\boldmathfont}{cmbx10}
\newfont{\headfontb}{cmbx10 scaled 1728}
\renewcommand{\theequation}{{\rm\thesection.\arabic{equation}}}
\renewcommand{\thefootnote}{\arabic{footnote}}

\vspace{1cm}
\begin{titlepage}
\begin{flushright}
NORDITA-2019-062
\end{flushright}
\vskip 2cm
\begin{center}
{\huge{\bf The Holographic QCD Axion}}
\end{center}

\vskip 10pt
\begin{center}
{\large
Francesco Bigazzi$^{1}$, Alessio Caddeo$^{1,2}$, Aldo L. Cotrone$^{1,2}$,\\[11pt]Paolo Di Vecchia$^{3,4}$, Andrea Marzolla$^{1,2,5}$
}
\end{center}
\vskip 10pt
\begin{center}
\vspace{0.2cm}
\textit {$^1$ INFN, Sezione di Firenze; Via G. Sansone 1; I-50019 Sesto Fiorentino (Firenze), Italy.}\\
\textit{$^2$ Dipartimento di Fisica e Astronomia, Universit\'a di Firenze; Via G. Sansone 1;\\ I-50019 Sesto Fiorentino (Firenze), Italy.}\\
\textit{$^3$ The Niels Bohr Institute, University of Copenhagen, Blegdamsvej 17,\\DK-2100 Copenhagen, Denmark.}\\
\textit{$^4$ Nordita, KTH Royal Institute of Technology and Stockholm University,\\Roslagstullsbacken 23, SE-10691 Stockholm, Sweden.}\\
\textit{$^5$ Centro At\'omico Bariloche and CONICET S.C. de Bariloche, R\'io Negro,\\ R8402AGP, Argentina.}\\

\vskip 20pt
{\scriptsize{bigazzi@fi.infn.it, alessio.caddeo@unifi.it, cotrone@fi.infn.it, divecchi@nbi.dk, andrea.marzolla@cab.cnea.gov.ar
}}

\end{center}

\vspace{25pt}

\begin{center}
 \textbf{Abstract}
\end{center}

\noindent 

A holographic model of QCD axion is presented.
It describes a composite axion in the KSVZ class.
Having a gravity dual, based on the Witten-Sakai-Sugimoto model, it is calculable in the strongly coupled regime.
Its basic properties are derived, including the low energy Lagrangian, from which the axion couplings to nucleons can be derived.
Basic features in the deconfined phase are studied as well.
In particular, the temperature dependence of the axion mass is extracted from the topological susceptibility. As an aside, the topological susceptibility of strongly coupled ${\cal N}=4$ SYM at finite temperature is derived for the first time.

\end{titlepage}
\newpage
\tableofcontents

\section{Introduction}

The QCD axion scenario provides an elegant solution to two distinct problems in particle physics: the strong CP problem and the nature of dark matter.
The strong CP problem concerns the reason why the QCD $\theta$-angle, {\it i.e.} the coefficient of the $F \wedge F$ term in the Lagrangian, is experimentally very small, of order $10^{-10}$, instead of being a number of order one.
The Peccei-Quinn mechanism relates this constant to the expectation value of a field, whose potential has a minimum at (or very close to) zero \cite{Peccei:1977hh,Wilczek:1977pj}.
The low energy excitations of this field comprise a scalar particle, the axion, which is a viable candidate for being a component of dark matter\cite{Preskill:1982cy}, provided the axion decay constant value $f_a$ falls in a specific energy range (see e.g.~the review \cite{Irastorza:2018dyq}).\footnote{
	The range of $f_a$ goes from $10^8$-$10^9$ GeV up to around $10^{17}$ GeV, see e.g.~\cite{villadoro}.}

Trading the fine-tuning of the strong CP problem for this ``fine tuning'' of $f_a$ seems a convenient deal, for it provides at the same time a dark matter candidate.
For this reason, in the last years there has been a renewed interest in building concrete axion models of different types and exploring their phenomenological consequences.
Moreover, the study of axion-like particles is theoretically interesting, since they are very common in top-down models of particle physics, for example they are ubiquitous in string compactifications \cite{wittenaxion}.

In this paper we show how to construct the first (to our knowledge) top-down holographic model of an axion whose low energy effective Lagrangian coincides with the axion-dressed chiral QCD one (see e.g. \cite{sannino,DiVecchia:2017xpu}), in the planar limit.
The model is built as a simple extension of the Witten-Sakai-Sugimoto (WSS) dual of a planar QCD-like theory \cite{Witten:1998zw,Sakai:2004cn}.
It is a composite axion model falling in the KSVZ class \cite{KSVZ}.
Its rationale is the following.

It is well known that a single massless quark is sufficient to solve the strong CP problem, since a massless fundamental fermion renders non-physical the $\theta$-angle.
Nevertheless, experiments are not consistent with a massless {\it up} quark and any additional massless quark flavor (a left and right quark pair) would be incompatible with phenomenology.

This problem can be avoided if the extra quark flavor condenses at a scale $M_a$ which is much higher than the ordinary chiral symmetry breaking scale ($f_a$ will be of the same order of $M_a$ in the model).
In this case the extra quarks do not enter perturbative corrections of Standard Model observables and all the extra hadrons associated to this flavor have masses of order $M_a$, so they are not excluded by phenomenology for sufficiently large $M_a$.
The one exception is the (pseudo) Goldstone boson of the extra chiral symmetry breaking.
This is a natural candidate for the QCD axion, since it solves automatically the CP problem by the standard Peccei-Quinn mechanism.
The axial $\mathrm{U}(1)_\mathrm{A}$ symmetry of the extra flavor plays the role of the broken symmetry involved in the Peccei-Quinn mechanism.

Obviously, the main point of this scenario is how to induce a condensation of the extra flavor at the scale $M_a$.
In the model presented in this paper, the condensation is induced by a strongly coupled version of a (non local) Nambu-Jona-Lasinio~(NJL) quartic interaction between the extra quarks.    
The NJL condensation scale $M_a$ is a genuine parameter of the model and can be made parametrically large.
In fact, already the standard NJL interaction induces chiral symmetry breaking without confinement, so that the two scales can be well separated.\footnote{Notice that the composite axion in our model is thus not a pseudo-Goldstone boson of some extra, hidden, gauge theory. In this respect the model differs from both standard composite axion models and from more recent scenarios based on a large $N$ hidden sector \cite{eliasbianchi}.}  

The quartic interaction requires a suitable UV completion.
In the holographic model at hand, this is automatically provided by a higher dimensional theory.
The whole QCD-like theory with axion, including its four-dimensional low energy phase, 
has, at strong coupling and in the planar limit, a dual gravitational description.
As such, 
the construction provides a concrete, calculable strongly coupled model for a composite axion. 
Needless to say, other UV embeddings of the quartic NJL operator, possibly not related to holographic models, are an interesting venue to explore in their own, but we will not pursue this issue in the present paper.

The holographic WSS model we exploit is recognized as a close cousin of QCD, sharing its low energy pattern of symmetry breaking.
It encompasses in an impressively precise way the low energy Chiral Lagrangian, the Skyrme model and their effective extensions including the vector and axial vector mesons. 
The model can be used to estimate various QCD observables with a precision which is often comparable to the one of other effective field theories. 

The building blocks of the WSS model are a gravity background in type~IIA supergravity, dual to a planar Yang-Mills theory, and a set of probe $D8$-branes supporting flavor degrees of freedom.
The geometry includes a cigar part and the two branches of the $D8$-branes are placed at antipodal points on the cigar circle.
These $D8$-branes are dual to the QCD flavors. Their current algebra masses can be induced by stringy instanton effects as we will review in the following. 
It turns out that adding a single $D8$-brane in a non-antipodal configuration provides the extra flavor (with zero current algebra mass) needed to implement the composite axion model described above.

In section \ref{secaxion} we describe more precisely this construction and present the low energy effective theory, discussing its parameter dependence and its main characteristics (technical details are presented in appendices \ref{appreview} and \ref{app1}).
In the confined phase, the model can be employed for instance to calculate the magnitude of the axion couplings to nucleons, which constitute important data for axion detection and cosmology.
The derivative axion couplings to the proton and neutron can be extracted from existing computations in \cite{hss}.
The non-derivative couplings have been derived in \cite{axioncouplings} for both the holographic and the Skyrme model. For completeness, we review these results in section \ref{sectionmatter}.

Then, in section \ref{secthermo}, we consider the axion model in the deconfined phase, focusing, for simplicity, on the Yang-Mills-like theory without probe QCD $D8$-branes. We address the temperature dependence of the axion mass, due to its crucial role in determining the relative axion/dark matter relic abundance in the cosmological evolution. As a first step, we compute the topological susceptibility of the Yang-Mills-like theory in the deconfined phase. The susceptibility is exponentially suppressed by the large number of colors. Its leading contribution is due to instantons and can be extracted from the known $D0$-instanton terms in type IIA string theory. As a second step, we also calculate the temperature dependence of the axion decay constant and from these data we extract the axion mass. As a bonus, we provide for the first time the result for the topological susceptibility of ${\cal N}=4$ SYM at finite temperature.

Since holographic models in the gravity regime do not describe the asymptotically free phase of the dual gauge theories, the temperature dependence of the axion mass turns out to depart decidedly from the one expected in QCD at large temperatures, where the dilute instanton gas approximation is under control.
Recent lattice studies seem to point towards the extension of the same ``large temperature'' behavior up to small temperatures, very close to the critical one \cite{Borsanyi:2016ksw}.\footnote{For recent computations of the topological susceptibility at finite temperature on the lattice see also \cite{Berkowitz:2015aua,Kitano:2015fla,Borsanyi:2015cka,Bonati:2015vqz,Petreczky:2016vrs,Frison:2016vuc,Burger:2018fvb,Bonati:2018blm,Giusti:2018cmp}.}
Thus, it is possible that the behavior of the actual QCD axion mass has no regime comparable to the holographic one, which is found to increase with the temperature.

We conclude the paper with a few comments in section \ref{seccomments}.
In appendix \ref{appendix:njl} we report some considerations on the NJL model in a theory with an extra dimension. 
 
\section{The holographic axion}
\label{secaxion}
\setcounter{equation}{0}
In this section we first briefly review the Witten-Sakai-Sugimoto model of holographic QCD and then describe the addition of the axion degree of freedom.
We derive the full low energy action, describe the axion main characteristics and present its couplings to the nucleons.

The building blocks of the WSS holographic QCD model are two different classes of \mbox{$D$-branes} in type IIA string theory. The non-supersymmetric Yang-Mills sector is realized as the low energy dynamics of $N_c$ $D4$-branes wrapped on a circle ($S_{x_4}$) where anti-periodic boundary conditions for the fermions are imposed. The chiral quark matter fields are the low energy modes of open strings stretching between the $D4$-branes and $N_f$ $D8/\overline{D8}$-branes placed at different points on the circle.   

In the 't Hooft large $N_c$, fixed $N_f$ limit at strong coupling, the holographic dual description of the model is provided by the near horizon limit of the gravity background sourced by the $N_c$ $D4$-branes, probed by the $D8$-branes. 

The background, found by Witten in \cite{Witten:1998zw}, actually provides a holographic dual description for a $\mathrm{SU}(N_c)$ Yang-Mills theory in $3+1$ dimensional Minkowski spacetime, coupled to massive Kaluza-Klein (KK) adjoint matter at low energies. We will often refer to it as the Witten-Yang-Mills (WYM) background. It contains a non trivial metric, a running dilaton and a four-form Ramond-Ramond (RR) field strength
\bea \label{wym}
ds^2&=&\Big(\bfrac{u}{R}\Big)^{3/2} \left(dx^\mu dx_\mu + f(u)d x_4^2 \right)+ \Big(\bfrac{R}{u}\Big)^{3/2}{du^2\over f(u)}
+R^{3/2}u^{1/2}d\Omega_4^2\ ,\nonumber \\     f(u)&=&1-{u_0^3\over u^3}\ ,  \qquad     e^\ff =g_s{u^{3/4}\over R^{3/4}}\ , \qquad F_4 = 3 R^3 \omega_4 \,.
\eea
Here $\mu=(0,1,2,3)$ label the 4d Minkowski directions, the radial variable $u$ has dimensions of length and ranges in $[u_0,\infty)$, $R=(\pi g_s N_c)^{1/3}l_s$ and $\omega_4$ is the volume form of the transverse~$S^4$. The circle wrapped by the $D4$-branes is parameterized by the compact coordinate $x_4$ which is thus taken to have finite extension~$\beta_4= 2\pi R_4 = 2\pi/M_\mathrm{KK}$. The adjoint Kaluza-Klein modes related to compactification have thus masses of order~$M_\mathrm{KK}$. 

Absence of conical singularities at $u\!=\!u_0$ is ensured setting $9 u_0 \beta_4^2\!=\!16 \pi^2 R^3$. With this condition the $(x_4, u)$ subspace has the topology of a cigar, a key feature of the model. Confinement and the formation of a mass gap for the glueballs are beautifully encoded by this holographic model. The glueball masses and thus the scale $\Lambda_\mathrm{YM}$, turn out to be proportional to $M_\mathrm{KK}$. The Yang-Mills string tension scales as $T_s\sim \lambda M_\mathrm{KK}^2$, where the parameter \mbox{$\lambda\!=\!2\pi g_s N_c l_s M_\mathrm{KK}\equiv g_{\mathrm{YM}}^2 N_c/2$} -- a sort of  't Hooft coupling at the scale $M_\mathrm{KK}$ -- measures how much the model departs from planar Yang-Mills. The holographic approximation is reliable only when $\lambda\gg1$, which implies that in such regime the spurious KK massive adjoint modes are never decoupled.  

A non zero topological $\theta$-angle in the model can be included by turning on a RR one-form potential $C_1$ along the circle $S_{x_4}$, so that, at $u\rightarrow\infty$, $\int_{x_4}C_1\sim \theta$ \cite{Witten:1998uka}. On the confining background (\ref{wym}) this boundary condition and the equations of motion give rise to a non trivial field strength $F_2=dC_1$ whose flux along the cigar is $\int_{(x_4,u)} F_2 \sim \theta$. 
If $\theta/N_c\rightarrow0$ the backreaction of this field on the background can be neglected. \footnote{It is also possible to take the backreaction into account, see \cite{barbon,dubovsky,wymtheta}.} Further details can be found in appendix \ref{appreview}.

The introduction of $N_f$ probe $D8/\overline{D8}$-branes at antipodal points on the circle $S_{x_4}$, corresponds to adding chiral quarks to the theory, as shown by Sakai and Sugimoto~\cite{Sakai:2004cn}. In the holographic background the two branches actually join at the tip of the cigar, nicely accounting for the spontaneous chiral symmetry breaking in the quantum field theory. The low energy effective action on the $N_f$ $D8$-branes precisely reproduces the 4d chiral Lagrangian for the pseudoscalars including the Witten-Veneziano mass term for the $\etaprime$. The tower of (axial) vector mesons is accounted for as well. 

Scalar and (axial) vector mesons in the model correspond to fluctuations of the $\mathrm{U}(N_f)$ gauge field $\cal A$ on the $D8$-branes.  
In particular, the matrix $U$ of the pseudoscalar bosons is given by the path-ordered holonomy $U= \mathcal{P} e^{i \int {\cal A}_z dz}$, where $z\in(-\infty,\infty)$ is a suitably, single-valued redefinition of the holographic radial variable on the $D8$-branes.  

A (small w.r.t.~$M_\mathrm{KK}$) current algebra mass term for the quarks and hence a term of the form $\Tr[M U +\text{h.c.}]$ in the effective action for the pseudoscalar fields, can also be added in the holographic model. All these features are reviewed in appendix~\ref{appreview}. 

The antipodal embedding is chosen in such a way that chiral symmetry breaking and confinement occur at the same energy scale.
This is not the only possibility.
One can generically place the $D8/\overline{D8}$-branes at a separation distance $L < \pi R_4$, where $R_4$ is the radius of the circle \cite{Aharony:2006da}. In this case the brane and the anti-brane join at a radial position $u_J >u_0$. Crucially, as shown in \cite{Antonyan:2006vw}, this deformation of the theory does not correspond to giving a current algebra mass to the quarks -- the pseudoscalar Goldstone bosons are still exactly massless \cite{Aharony:2006da}. Rather, the new parameter $L$ (or equivalently $u_J$) is dual to the coefficient of a deformation of the theory which, at weak coupling, is a non-local quartic coupling between the quarks. As such, this type of embedding is suitable to study the strongly coupled phase of a NJL-like model. 

As we discuss in detail in appendix \ref{appendix:njl}, the non-locality of the quartic term at weak coupling depends on the fact that the fermions are stuck at different points in the fifth dimension. Once the 5D gauge field is integrated out (in the 1-gluon exchange approximation) the quartic terms arise. It is worth to outline that this perturbative picture does not hold when the dual holographic description is reliable. 
\begin{figure}[ht!]
\centering
\includegraphics[scale=.5]{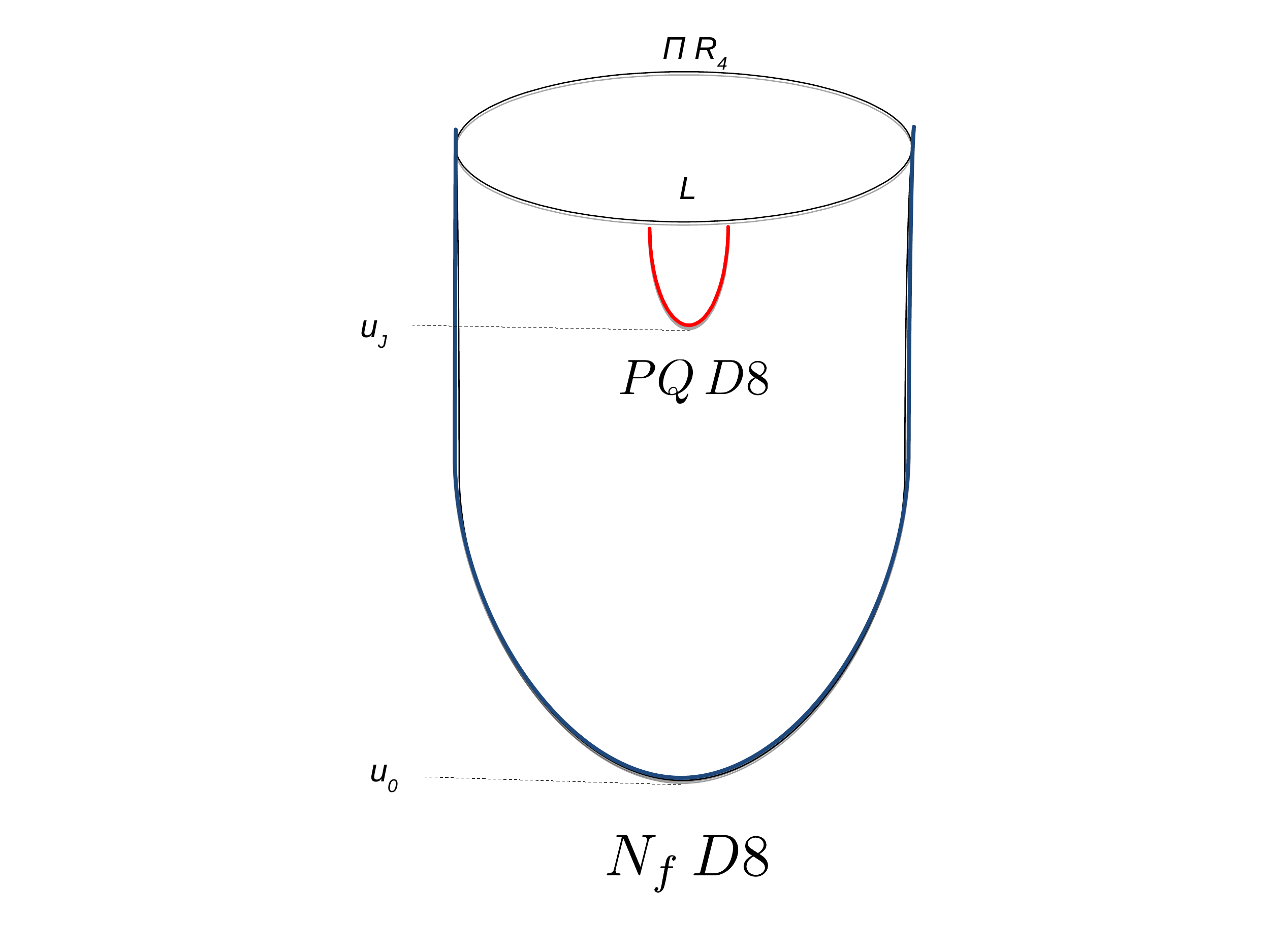}
\caption{Brane arrangement for the holographic axion model.}
\label{ssa}
\end{figure}

We are going to point out that simply adding to the WSS model with antipodally embedded $D8$-branes, one extra $D8/\overline{D8}$-brane pair (a ``PQ-brane'', where PQ stands for Peccei-Quinn) with $L < \pi R_4$ {\it corresponds to adding an axion} to the QCD-like theory (see figure~\ref{ssa}).   
In fact, as we have seen, the addition of this $D8/\overline{D8}$ pair corresponds to the addition of an extra flavor of chiral quarks with a quartic Nambu-Jona-Lasinio type interaction.\footnote{An axion model with some similarities in a non holographic context can be found for instance in~\cite{shiu}. A five dimensional orbifold model where the QCD axion is identified with the component of a $U(1)$ gauge field along the extra dimension, has been proposed in \cite{choi}.}
As stressed above, these quarks are still massless, thus accounting for the Peccei-Quinn mechanism trivializing the $\theta$ angle, much like a massless $up$ quark would do. We know that a massless $up$ quark is excluded by experiments. In the case at hand, instead, the extra flavor condenses due to the quartic coupling, at a scale dictated by the quartic coupling itself. The latter is a free parameter of the theory, so the condensation scale can be made very high. In this way the only degree of freedom surviving at low energy is the (would be) Goldstone boson of the symmetry breaking, the axion $a$.
All the other mesons have a very large mass of the order of the condensation scale, both the ones with two PQ valence quarks \cite{Aharony:2006da} and the ones with one PQ and one QCD valence quark \cite{heavylight}.

The low energy physics of the axion is dictated by symmetry: it couples in the Chiral Lagrangian just to the topological charge density operator $Q \sim \Tr F \wedge F$, precisely as the $\etaprime$.
But, crucially, its decay constant $f_a$ is not related to $f_{\pi}$ (not even in the planar limit we are discussing, where $f_{\eta'} =f_{\pi}$), since it is controlled by the new free parameter $L$.

In the following subsection we will present the effective action of the model, focusing on the axion plus pseudoscalar mesons sector. We will then discuss the axion couplings to matter. 

In appendix \ref{appendix:njl} we will extend some of the weak coupling considerations of \cite{Antonyan:2006vw} on the non-local NJL model to the case with two types of flavors. The reliability of the calculation in the appendix requires that the separations of both flavors, $L$ for the PQ quarks and $L'$ for the QCD ones, is much smaller than $\pi R_4$.  This is different from the WSS antipodal model (where the QCD quarks are at distance $\pi R_4$), but allows to have an idea on the kind of quartic operators which are turned on, and the relative strength of their couplings. 
In particular, it is found that, apart from the two quartic terms involving separately the QCD and PQ quarks, there are quartic terms involving both types of quarks.
The strength of the coupling of the latter interaction is of the same order of the one among the QCD quarks, so it is supposed to be much smaller than the one of the PQ quarks in the limit $L \ll L'$ we are interested in. 
So, the dominant term at weak coupling is the one causing the condensation of the PQ quarks at the scale $f_a$.

\subsection{The effective action}
 
To lowest order in derivatives, the 4d low energy effective Lagrangian on the $N_f$ antipodal WSS $D8$-branes plus the extra non-antipodal PQ one, reduces to 
\be
{\cal L}_\text{eff} = - \frac{f_{\pi}^2}{4}\Tr\big[\partial_{\mu}U \partial^{\mu}U^{\dagger}\big] -\bfrac12 \partial_{\mu}a\partial^{\mu}a + c \Tr\big[M U^\dagger +\text{h.c.}\big] -\frac{\chi_\mathrm{WYM}}{2}\bigg(\theta + \frac{\sqrt{2N_f}}{f_{\pi}}\etaprime + \frac{\sqrt{2}}{f_{a}}a\bigg)^{\!2}\ ,
\label{leffe}
\ee
where $a$ is the axion field, $U = e^{2i (\Pi^a T^a + (2N_f)^{-1/2} \etaprime ) / f_{\pi}}$ is the $\mathrm{U}(N_f)$ matrix for the pseudoscalar bosons $\Pi^a$ and $\etaprime$, $T^a$ are $\mathrm{SU}(N_f)$ generators (with $\Tr [T^a T^b]=\delta^{ab}/2$), $f_{\pi}$ (resp.~$f_a$) is the pion (resp.~axion) decay constant and we set $f_{\pi}=f_{\etaprime}$ since we work in the large $N_c$ limit and with small quark masses. 
Finally, $M$ is the QCD flavor mass matrix, $c$ is proportional to the VEV of the chiral condensate, ($-2c = \langle\bar{q}_i q_i \rangle$ that, in the large $N_c$ limit and for small quark masses, does not depend on the flavor index $i$) and $\chi_\mathrm{WYM}$ is the topological susceptibility of the unflavored theory. 
To this Lagrangian one can add systematically other pieces, such as the Skyrme term and (axial) vector meson contributions, see appendix \ref{appreview}. 
Notice that the physical axion, $\etaprime$ and pions will actually be given by linear combinations of the fields appearing above, as it can be easily realized by diagonalizing the related mass matrix. We will discuss this point in the next subsection.

When no axion field is included, the above effective Lagrangian is deduced from the low energy dynamics of the $N_f$ WSS $D8$-branes \cite{Sakai:2004cn,Bartolini:2016dbk}. We will briefly review the construction in appendix \ref{appreview}. In terms of the model parameters, we get
\be
f_{\pi}^2 = \frac{N_c \lambda}{54\pi^4}M_\mathrm{KK}^2\,,\quad \chi_\mathrm{WYM} =\frac{\lambda^3 M_\mathrm{KK}^4}{4(3\pi)^6}\,.
\label{paraWSS}
\ee
The connection between the masses of the quark flavors  $m_i$ entering in the matrix $M$ and the squared masses of the meson flavors  $\mu_i^2$ is given by the GMOR relation:
\be
-2m_i \langle {\bar{\psi}}  \psi \rangle = f_\pi^2 \mu_{i}^2 \,. 
\label{GMOR}
\ee
Analogously, as we will review in appendix \ref{app1}, in the unflavored model without the $N_f$ Sakai-Sugimoto branes, the low energy effective action for the axion field $a$ arises from the low energy limit of the non-antipodal PQ-brane.

Crucially, the interaction between the low energy modes on the WSS and PQ $D8$-branes is driven by the potential term proportional to $\chi_\mathrm{WYM}$ in (\ref{leffe}). It arises from the on-shell ten-dimensional action term $\int d^{10}x \,\big|\tilde F_2\big|^2$, where $\star \tilde F_2 = dC_7$. The modified RR field strength $\tilde F_2$ reduces to the standard $F_2=dC_1$ one in the pure Yang-Mills case where, as we have recalled above, $\int_{(x_4,u)} F_2\sim \theta$. In this case, the related on-shell action term $\int d^{10}x \big|F_2\big|^2$ provides the ${\cal O}(\theta^2)$ correction to the gauge theory energy density, which is in fact proportional to the topological susceptibility $\chi_\mathrm{WYM}$ \cite{Witten:1998uka}. In presence of $D8$-brane sources this term gets modified. 

The $N_f$ WSS antipodal $D8$-branes and the extra PQ non-antipodal one, in fact, carry CS action terms of the 
form\footnote{In this discussion we use labels in order to distinguish gauge fields and field strength belonging to the PQ-brane from those belonging to the WSS-ones. 
In the rest of the paper we will omit these labels since there will be no risk of confusion.}
$\int_\mathrm{WSS} C_7\wedge \Tr {\cal F} ^{\mathrm{WSS}}$ and $\int_\mathrm{PQ} C_7\wedge {\cal F} ^{\mathrm{PQ}}$, 
where $\mathcal{F} ^{\mathrm{WSS}}$ (resp.~$\mathcal{F} ^{\mathrm{PQ}}$) is the $\mathrm{U}(N_f)$ (resp.~$\mathrm{U}(1)$) gauge field strength on the 
WSS (resp.~PQ) branes.  These terms imply that the Bianchi identity for $\tilde F_2 \sim \star dC_7$ is modified, in such a way that 
$\tilde F_2 = dC_1 + \Tr{\cal A}^{\mathrm{WSS}}\wedge\omega_\mathrm{WSS} + A^{\mathrm{PQ}} \wedge \omega_\mathrm{PQ}$, where the one-forms  $\omega_\mathrm{WSS}$ and $\omega_\mathrm{PQ}$ are related to the $D8$-branes embeddings. 
Integrating the previous expression along the cigar, recalling that $\etaprime/f_{\pi}\sim \int dz \Tr{\cal A}^{\mathrm{WSS}}_z$ and 
identifying the novel pseudo-Goldstone boson as $a/f_a\sim \int A^{\mathrm{PQ}}_z$ one gets $\int\tilde F_2\sim \theta + \sqrt{2N_f}\etaprime/f_{\pi} + \sqrt{2}a/f_a$. 
The $\tilde F_2$ on-shell action term thus gives the whole potential term proportional to $\chi_\mathrm{WYM}$ appearing in (\ref{leffe}).  
In the standard WSS case ({\it i.e.} with no PQ $D8$-brane), this is precisely what produces the Witten-Veneziano 
(WV) mass term for the $\etaprime$ field \cite{Sakai:2004cn}. See appendix \ref{appreview} for further details.

When $N_f=0$, the above mentioned potential term in~\eqref{leffe} implies that the axion mass is given by the WV formula 
\be
m_{a}^2 = \frac{2}{f_{a}^2}\chi_\mathrm{WYM}\,.
\label{mWVaxion}
\ee	
In the flavored case, with massive flavors, $\chi_\mathrm{WYM}$ is replaced by the full topological susceptibility. We will briefly discuss the mass spectrum in subsection \ref{sectionmatter}.

All in all, by including both the $N_f$ antipodal $D8/\overline{D8}$ pairs and the non-antipodal PQ one, we get precisely the Chiral Lagrangian of~\cite{DiVecchia:2017xpu}. 
In order to see this explicitly, one can introduce an auxiliary field $Q(x)$ (that turns out to be the topological charge
density) and rewrite the last term in (\ref{leffe}) as follows
\bea
&& -\left(\theta + \frac{\sqrt{2N_f}}{f_{\pi}}\etaprime + \frac{\sqrt{2}}{f_{a}}a\right) Q + \frac{1}{2\chi_\mathrm{WYM}}Q^2 = \nonumber\\
&& =\left[-\theta +\frac{i}{2}{\rm Tr}\left(\log{U} -\log{U^{\dagger}} \right)  +\frac{i}{2}\left(\log{V} -\log{V^{\dagger}} \right) \right] Q + \frac{1}{2\chi_\mathrm{WYM}}Q^2 \,,
\eea
with $V=e^{i\frac{\sqrt{2}}{f_a}a}$ (this is called $N$ in \cite{DiVecchia:2017xpu}).

As we will show in appendix \ref{app1}, where details on the non-antipodally embedded flavor branes and the related meson spectrum are provided, the axion decay constant is given by
\be	\label{fazeroT}
f_a ^2 = \frac{N_c \l}{16 \pi^3} \frac{J^3(b)}{I(b)}  \frac{1}{M_\mathrm{KK} L^3} \,,\q \quad b\equiv \frac{u_0}{u_J}\ ,
\ee
where
\be
J(b) = \frac{2}{3} \sqrt{1-b^3} \int_0 ^1 dy \frac{y^{\frac12}}{(1-b^3 y)\sqrt{1-b^3 y -(1-b^3)y^{\frac83}\,}}\ ,
\ee
\be
I(b) = \int_0 ^1\!dy\ \frac{y^{-\frac12}}{\sqrt{1-b^3y -(1-b^3)y^{\frac83}\,}} \ ,
\ee
and
\be
\label{f1}
L = J(b) R^{\frac32} u_J ^{-\frac12}\,
\ee
is the distance between the PQ asymptotic branches on the $S_{x_4}$ circle.

Let us complete our analysis with the implications for our model coming from the phenomenological constraints 
which read 
\be
10^{9} \lesssim r \equiv \frac{f_a}{f_\pi}  \lesssim 10^{18}  \ .
\ee
Recalling eq.~(\ref{paraWSS}) for $f_{\pi}$, we get
\be
\label{conditiononmkkL}
M_\mathrm{KK} L = \bfrac{3}{2} \pi^{\frac13} J(b) I(b) ^{-\frac13}  r^{-\frac23} \ .
\ee
This condition fixes the value of $L$ which, in turn, fixes $u_J$. 
Using (\ref{f1}) and $u_0 = \tfrac49 M_\mathrm{KK} ^2 R^{3}$ this condition becomes
\be
b = \bigg(\frac{\pi}{r^2I(b)}\bigg)^{\!\frac23}\ .
\ee
Since $I(b) \simeq 2.4$ for any value of $b < 10^{-1}$, we obtain 
\be
10^{-24} \lesssim b \lesssim 10^{-12} \ .
\ee
Throughout this interval numerically $J\pr{b} \simeq 0.7$, so $L \simeq 0.7 \, R^{3/2}u_J ^{-1/2}$ and
\be
f_a^2 \approx 0.1534 \frac{N_c \l}{16 \pi^3 } \frac{1}{M_\mathrm{KK} L^3}\,.
\label{faT0}
\ee
We can compare the separation of the PQ-branes with the length of the cigar circle $\b_4$.
Recalling $\b_4 = 4 R^{3/2} u_0 ^{-1/2} /3$ we have $L/\b_4 \simeq 0.5 \, b^{1/2}$ and therefore
\be
10^{-12} \lesssim \frac{L}{\b_4} \lesssim 10^{-6} \ .
\ee
We know that our model is reliable in a range of the radial coordinate below a critical value $u_{\text{crit}}$, which can be identified 
as that value for which the dilaton $e^\ff$ becomes of order one. Recalling (\ref{wym}), this value is given by
\be
u_{\text{crit}} \sim R g_s ^{-4/3}  \sim \frac{N_c ^{4/3} l_s ^2 M_\mathrm{KK}}{\l} \ .
\ee
We therefore ask the condition $u_J \ll u_{\text{crit}}$ to be valid. Recalling $2R^3 l_s ^{-2} = \l/M_\mathrm{KK}$, we can write 
this condition in terms of field theory quantities as
\be
N_c^{2/3} \gg r^{2/3}  \l \ .
\ee

\subsection{Further properties of the model}

In the setting outlined above, the PQ-brane is put at a macroscopic distance in the internal space from the $N_f$ WSS branes.
As a consequence, in the dual theory the two types of degrees of freedom interact only through the gauge sector.
Since we are in the planar limit, these interactions are quite suppressed.
Thus, this model falls in the KSVZ class \cite{KSVZ} (the PQ symmetry is not realized on Standard Model fermions).
As such, the UV dependent part - {\it i.e.} that not encoded in the low energy Lagrangian \eqref{leffe} - of the couplings of the axion with the nucleons is exactly zero, as reviewed for instance in~\cite{villadoro}.
In the next subsection we will collect the known results for the ``universal'', IR part of these couplings \cite{axioncouplings}. 

Moreover, in the WSS model the electromagnetic current is obtained by weakly gauging a vectorial part of the $\mathrm{U}(N_f)\times \mathrm{U}(N_f)$ chiral symmetry. As such, it pertains to the $N_f$ WSS $D$-branes: the PQ quarks are 
automatically electrically uncharged.\footnote{Strictly speaking, in the holographic model, the electromagnetic $U(1)$ symmetry is global on the QFT side, since it corresponds to a gauge symmetry in the bulk. It is not clear whether our PQ quarks would stay uncharged also in a setup where a local $U(1)$ Maxwell symmetry is properly realized in the QFT. We thank Michele Redi for discussions about this point.}   

Thus, the electromagnetic interactions of the axion just come from the mixing with the pion and the $\etaprime$ in the Chiral Lagrangian, with no UV contribution. Also in this respect, the model falls in the KSVZ class.
Apart from the existence of a strongly coupled version of the NJL interaction in the UV, the extra information that the holographic picture seems to provide about the model is the co-existence of (at least) three quark condensates in the IR.
In fact, suspended between each type of $D8$-branes there are string world-sheets whose areas provide the magnitude of the fermion bilinear condensates \cite{AK}. Hence, if we denote as $q, \psi$ respectively the standard and extra quarks we expect to have both the
$\langle \bar q q \rangle$, $\langle \bar \psi \psi \rangle$ condensates and a condensate of the form $\langle \bar q \psi \bar \psi q \rangle$. It would be interesting to investigate whether the latter has some influence on the model both at weak and at strong coupling. 

A last comment is in order.
We have presented the simplest axion model with one extra flavor and a very symmetric configuration of branes.
This is not the only possibility.
For example, one can imagine a setting where one end point of the PQ-brane coincides with one end point of the QCD-branes, so that one stack of branes contains $N_f+1$ elements. Nevertheless, there is no enhancement of chiral symmetry (before spontaneous chiral symmetry breaking). This is because the QCD quarks have a mass term breaking explicitly chiral symmetry to $\mathrm{SU}(N_f) \times \mathrm{U}(1)_{B} \times \mathrm{U}(1)_{L} \times \mathrm{U}(1)_{R}$, where the last two terms refer to the PQ quarks.

In the dual holographic picture, this means that the $N_f$ WSS $D8$-branes {\it must} join the $N_f$ WSS $\overline{D8}$-branes.
There is no possibility of joining one WSS $D8$ with the PQ $\overline{D8}$, for example.
Then, the PQ $D8$ and $\overline{D8}$ have no possibility but joining among themselves.

\subsection{Axion couplings to matter}\label{sectionmatter}
In order to be self-contained, in this subsection we report on the results found in \cite{axioncouplings} on the couplings of the axion to the 
nucleons\footnote{Notice that the axion decay constant $f_a$ in the present paper differs from that in \cite{axioncouplings}: 
$ f_a \pq{\rm{here}}= \sqrt{2} f_a [\rm{there}]$.}.
As we have stressed before, the UV part of these couplings is exactly zero.
The couplings can then be extracted directly from the low energy action (\ref{leffe}).

Let us consider the $N_f=2$ case, where $M={\rm diag}(m_u, m_d)$ and $2 T^a=\tau^a$ are the Pauli matrices. Neglecting the mixing terms between pions 
and $\etaprime$, the mass eigenstates of the Lagrangian (\ref{leffe}), to leading order in $1/f_a$ and absorbing the $\theta$ parameter in the VEV 
of $ \sqrt{2} a/f_a$, turn out to be
\bea\label{masseigen}
&&\hat \etaprime = \etaprime + \frac{\chi f_{\pi}}{4 c}\Tr [M^{-1}] \frac{\sqrt{2} a}{f_a}\,,\nonumber \\
&&\hat \pi^a = \pi^a + \frac{\chi f_{\pi}}{4 c}\Tr [\tau^a M^{-1}] \frac{\sqrt{2} a}{f_a}\,,\nonumber \\
&& \hat a = a - \frac{\chi f_{\pi}}{4 c}\Tr [M^{-1}] \frac{\sqrt{2} \etaprime}{f_a} - \frac{\chi f_{\pi}}{4 c}\Tr [\tau^a M^{-1}] \frac{\sqrt{2} \pi^a}{f_a}\,,
\eea
where
\be
\chi= \frac{4 c \chi_\mathrm{WYM}}{4 c + 2\chi_\mathrm{WYM} \Tr[M^{-1}]}\,,
\label{fulltopo} 
\ee
is the full topological susceptibility of the theory. 

As it is shown in \cite{axioncouplings}, the model allows to deduce the derivative as well as the non-derivative couplings of the axion to matter fields to leading order in the holographic limit. The most relevant ones are the couplings with nucleons. In the WSS model, the latter correspond to instanton solutions for the gauge field $\cal F$ on the WSS $D8$-branes \cite{hataSS}. By carefully taking into account the mixing terms in (\ref{masseigen}), the axion couplings to nucleons can be derived from those of the pseudoscalar mesons. 

The derivative axion-nucleon couplings defined by the effective interaction term
\be
\delta{\cal L}_{aNN,{\rm der}} = -\frac{\partial_{\mu} a}{\sqrt{2} f_a} c_N \bar N \gamma^{\mu}\gamma^5 N\,,
\label{effder} 
\ee
where $N=(p,n)$ is the nucleon field, are given by
\bea
&& c_p = -\frac{g_{\etaprime NN}}{m_N} \frac{\chi f_{\pi}}{4 c } \Tr[M^{-1}] - \frac{g_{\pi NN}}{m_N} \frac{\chi f_{\pi}}{4 c} \Tr[M^{-1}\tau^3]\,\nonumber \\
&& c_n = -\frac{g_{\etaprime NN}}{m_N} \frac{\chi f_{\pi}}{4 c} \Tr[M^{-1}] + \frac{g_{\pi NN}}{m_N} \frac{\chi f_{\pi}}{4 c} \Tr[M^{-1}\tau^3]\,,
\label{cpcn} \eea
where $m_N$ is the (large) nucleon mass and $g_{\etaprime NN}$, $g_{\pi NN}$ are the CP-even couplings of the $\etaprime$ and the pions with the nucleons. 

To leading order in the chiral limit $4 c \ll 2\chi_\mathrm{WYM} \Tr [M^{-1}]$, which sets the pion mass to be much smaller than that of the $\etaprime$, the above expressions reduce to
\bea
\label{cpcnbis}
&& c_p \approx - \frac{1}{2}\hat g_A - \frac{1}{2} g_A \frac{m_d - m_u}{m_d+m_u}\,,\nonumber \\
&& c_n \approx - \frac{1}{2}\hat g_A + \frac{1}{2} g_A \frac{m_d - m_u}{m_d+m_u}\,,
\eea
where we have used the generalized Goldberger-Treiman relations 
 \be
g_{\etaprime NN} = \frac{m_N}{f_{\pi}}\hat g_A\,,\quad g_{\pi NN} = \frac{m_N}{f_{\pi}}g_A\,,
\label{GT} 
\ee
between the isoscalar and isovector axial couplings and those of the $\etaprime$ and the pion. 
The couplings in (\ref{cpcnbis}) are independent on any UV parameter and precisely coincide with the ones obtained in any axion model in the KSVZ class \cite{KSVZ}. In the WSS model, the couplings $g_A$ and ${\hat g}_A$ have been computed in \cite{hss}. To leading order in the holographic limit (and so in the classical limit for the instanton pseudo-moduli) they read
\be
\hat g_A = \frac{27}{2\lambda}\,,\qquad g_A = \frac{2N_c}{\pi}\sqrt{\frac{2}{15}}\,.
\ee

Non-derivative axion-nucleon couplings of the form
\be
\delta{\cal L}_{aNN, \text{non-der}} =  {\bar c}_N a \bar N N\,,
\label{ndlag} 
\ee
can be induced whenever the Peccei-Quinn mechanism is not precise and an effective $\theta$ parameter is left over. In \cite{axioncouplings} it has been found that
\bea
&& \bar c_p = - \bar g_{\etaprime NN} \frac{\sqrt{2} \chi f_{\pi}}{4 c f_a} \Tr[M^{-1}] - \bar g_{\pi NN} \frac{\sqrt{2} \chi f_{\pi}}{4 c f_a} \Tr[M^{-1}\tau^3]\,\nonumber \\
&& \bar c_n =  - \bar g_{\etaprime NN} \frac{\sqrt{2} \chi f_{\pi}}{4 c f_a} \Tr[M^{-1}] + \bar g_{\pi NN} \frac{\sqrt{2} \chi f_{\pi}}{4 c f_a} \Tr[M^{-1}\tau^3]\,,
\label{barcpcn} 
\eea
where $\bar g_{\etaprime NN}$ and $\bar g_{\pi NN}$ are the linear in $\theta$ CP-odd couplings of $\etaprime$ and pions with nucleons.\footnote{See for example Ref.~\cite{sannino} for a review on such terms in the chiral effective field theory.}
To leading order in the holographic limit, in \cite{axioncouplings} it has been found that
\bea
\bar g_{\etaprime NN} &=& - \left(\frac{54}{125}\right)^{1/4}\frac{ m_{\pi}^2 N_c^{3/2} \gamma_1}{f_{\pi}^2 \pi^{7/2}} (1-\epsilon^2)\theta= - \frac{\delta M_N}{2 f_{\pi}}(1-\epsilon^2) \theta\,,\nonumber \\
\bar g_{\pi NN}&=&-\frac{9}{8}\left(\frac{3}{10}\right)^{1/4} \frac{m_{\pi}^2 \sqrt{N_c} \gamma_2}{8 f_{\pi}^2 \pi^{3/2} \lambda}(1-\epsilon^2)\theta= -\frac{(M_n - M_p)_{st.}}{4 f_{\pi} \epsilon}(1-\epsilon^2)\theta\,,
\label{gencp}
\eea
where $m_{\pi}^2 f_{\pi}^2=2c (m_u+m_d)$, $\epsilon\equiv(m_d-m_u)/(m_d+m_u)$ and
\bea
\gamma_1 &=& \int_0^\infty d y\, y^2\left(1+\cos \frac{\pi}{\sqrt{1+1/y^2}}\right)\approx 1.10\,,\nonumber \\
\gamma_2 &=& \int_0^{\infty} dy (1+y^{-2})^{-3/2} \sin\left(\frac{\pi}{\sqrt{1+y^{-2}}}\right) \approx 1.05\,.
\eea
In (\ref{gencp}), $\delta M_N$ is the quark mass contribution to the nucleon mass in the mass degenerate case - the so-called pion-nucleon sigma term - \cite{hashimoto:2009hj} and $(M_n - M_p)_{st.}$ is the strong force contribution, due to isospin-breaking, to the neutron-proton mass splitting \cite{bini}. 

Recalling that $f_{\pi}^2 \sim N_c$, we see that $\bar g_{\etaprime NN}\sim N_c^{1/2}$ and $\bar g_{\pi NN}\sim N_c^{-1/2}$. 
Numerical estimates and further details on the couplings can be found in \cite{axioncouplings}.

\section{The holographic axion at large temperature} \label{secthermo}
\setcounter{equation}{0}
Interesting questions concerning axion phenomenology are posed at finite temperature.
In the holographic model, turning on a temperature amounts to modify the dual gravity solutions in such a way that the Euclidean time coordinate is a circle of length $\beta=1/T$. In the WYM case there are two possible solutions which satisfy this requirement. A solution is given again by eq.~(\ref{wym}), just with a compact time coordinate. The other possible solution is given by a background where the dilaton and the RR four form are the same as in (\ref{wym}), while the metric has an event horizon. Explicitly, it reads
\bea \label{wymT}
ds^2&=&\left({u\over R}\right)^{3/2} \left(-\tilde f(u)dt^2 +dx^i dx_i + d x_4^2 \right)+ \left({R\over u}\right)^{3/2}{du^2\over \tilde f(u)}
+R^{3/2}u^{1/2}d\Omega_4^2\ ,\nonumber \\     \tilde f(u)&=&1-{u_T^3\over u^3}\ ,
\eea
with $i=1,2,3$. Here $u_T$ is the radial position of the horizon, related to the temperature by
 \be
u_T = \frac{16}{9}\pi^2 R^3 T^2\,.
\label{temp0}
\ee
This black brane solution, whose entropy scales like $N_c^2$, corresponds to the deconfined phase of the dual gauge theory and it is energetically preferred at $T>T_c$ where $T_c=M_\mathrm{KK}/(2\pi)$. The other solution mentioned above is dual to the confined phase at $0\!<\!T\!<\!T_c$. The holographic model precisely accounts for a first order phase transition at $T=T_c$ between the two phases.  

Notice that the temperature dependence in the confined phase is typically trivial, {\it i.e.} no dependence is found, basically because of the absence of a horizon. 
This means that the main properties of the holographic axion will not be sensible to the temperature as far as $T<T_c$.

Before going on let us just recall that the black brane solution presented above has, notoriously, two unwelcome features.
To begin with, it has been argued that the dual QFT phase is not in the same universality class of finite temperature Yang-Mills (or QCD for that matters)\footnote{Instead, the confined phase dual solution (\ref{wym}) is commonly believed to be in the same universality class of pure YM in the confined phase.} - some discrete symmetries do not match \cite{Mandal:2011ws}.\footnote{In \cite{Mandal:2011ws} there is an attempt to build the ``correct'' deconfined phase dual solution.}
Moreover, the deconfined phase dual solution reflects the six-dimensional nature of the holographic model in the UV; for example, the free energy density scales with the sixth power of the temperature.
Nevertheless, the solution is very simple and allows to get a lot of geometric intuition on chiral symmetry restoration and deconfinement transition, so we are going to use it as a proof of concept. 

\begin{figure}[ht!]
\centering
\includegraphics[scale=.5]{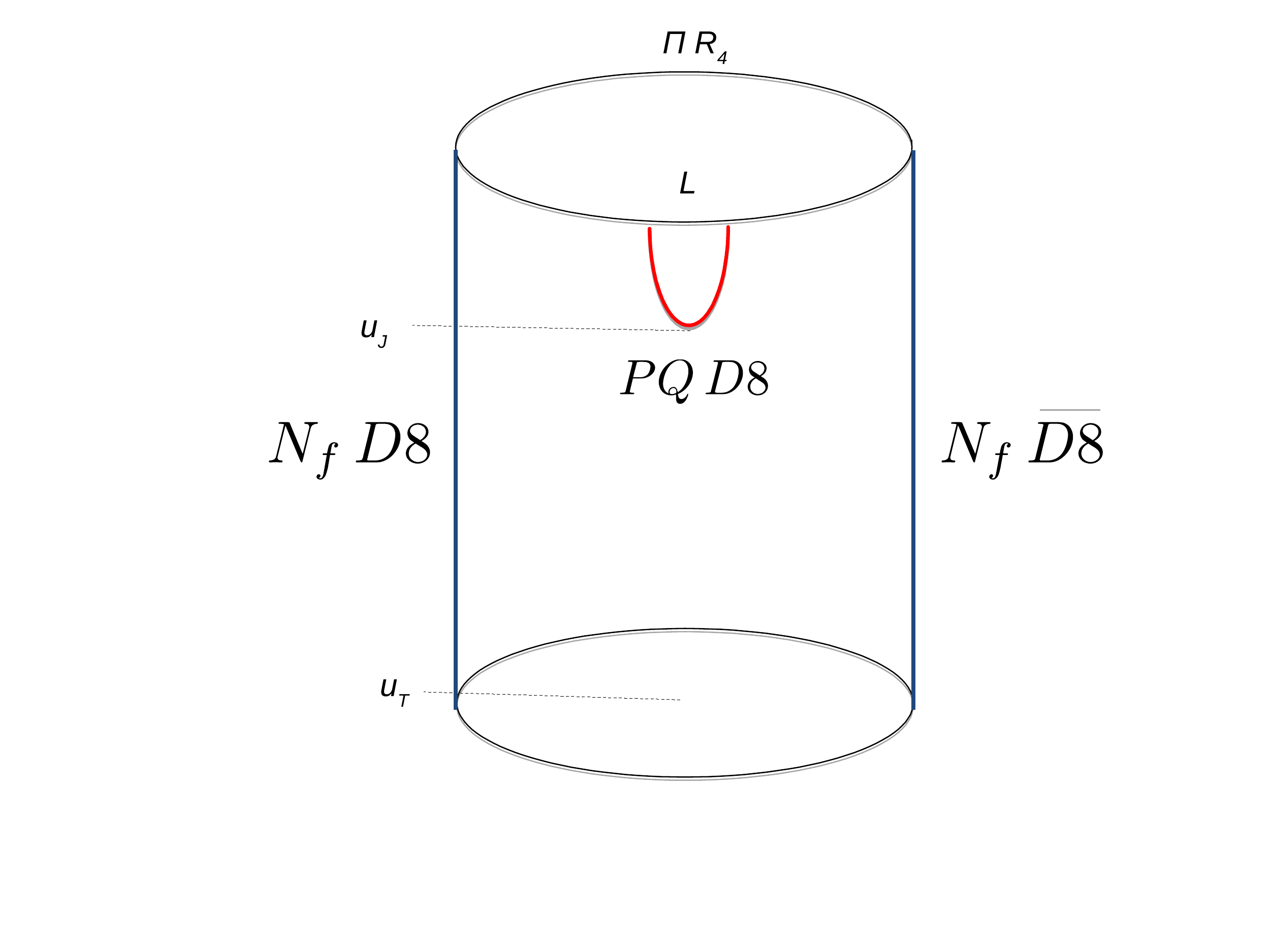}
\caption{Brane arrangement in the deconfined phase.}
\label{fig2}
\end{figure}

The main feature of the black brane solution (\ref{wymT}) is that it can be obtained by a double Wick rotation of the solution in (\ref{wym}), exchanging the role of~$S_{x_4}$ with the Euclidean temporal circle (with length $\beta \sim 1/T$).
As such, the cigar in the geometry (\ref{wym}) is replaced by a semi-infinite cylinder originating at $u_T$, see figure \ref{fig2}.

This makes it clear that the two branches of the $N_f$ WSS antipodal $D8$-branes fall separately into the horizon and do not join anymore, realizing chiral symmetry restoration in the deconfined phase. Instead, the PQ brane is quite mildly affected by the presence of the horizon as long as $u_T \ll u_J$, that is $T \ll {\cal O}(10^{10})$ GeV (at that temperature the axion would melt in the plasma, revealing its composite nature). 
This means that the embedding of the PQ brane will have some quite small deformation due to the (distant) horizon \cite{Aharony:2006da}.   

Concerning the axion, a first important issue we want to focus on is the temperature dependence of its mass in the deconfined phase.
The axion mass, in turn, is expected to be proportional to the topological susceptibility. In the flavored WSS setup we know that the latter can be different from zero only if all the quarks are massive. However, at present, there is no clear prescription on how to give finite masses to the flavors in the deconfined phase of the model. For this reason, we will only consider the unflavored WYM model in the following. 

The topological susceptibility of the WYM model at $T>T_c$  is certainly strongly suppressed in the planar limit. This is reflected by the fact that the solution for the RR one-form potential $C_1$, such that $\int_{S_{x_4}} C_1 \sim \theta$, is now a constant on the black brane background. Thus, $F_2=dC_1 =0$, so that at leading order in $N_c$ there are no effects of the $\theta$ angle in this phase. 

In order to recover a non-trivial $\theta$-dependence it is necessary to go beyond the leading order gravity approximation. In Yang-Mills we know that, at least for asymptotically large temperatures, the $\theta$-dependence is captured by a dilute instanton gas. The leading one-instanton contribution to the topological susceptibility gives $\chi_\mathrm{YM}(T)\sim T^4 e^{-8\pi^2/g_\mathrm{YM}^2(T)}$, where the gauge coupling is evaluated at the scale $T$. Asymptotic freedom drives the exponentially suppressed term into $\chi(T)\sim T^4 (T/\Lambda)^{-\frac{11}{3}N_c}$, which for $N_c=3$ gives a susceptibility which is power-like suppressed $\chi_\mathrm{YM}\sim T^{-7}$.  

In the holographic WYM model the one-instanton action corresponds to the action for a Euclidean $D0$-brane whose worldline is along the compact direction $x_4$.
 This brane is stable only in the deconfined phase, since in the confined case the cigar geometry tends to shrink the $D0$-brane to zero size. This nicely reproduces the field theory expectation that no (dilute) instanton gas can be defined in the confined phase.
The Euclidean $D0$-brane action on the background (\ref{wymT}) supported by a constant $C_1\sim\theta dx_4$ potential, reads
\be
S_{D0} = \bfrac{1}{l_s} \int e^{-\phi} \sqrt{g_{44}} dx_4 -\bfrac{i}{l_s}\int C_1 = \frac{8\pi^2}{g_\mathrm{YM}^2} - i \theta\,,
\label{d0act}
\ee
where $g_\mathrm{YM}^2\equiv 4\pi g_s l_s M_\mathrm{KK}$. We thus expect instanton driven exponential corrections $e^{-S_{D0}} + \text{c.c.}$ to contribute to the topological susceptibility. Remarkably enough string theory allows to precisely compute these corrections. As we will see in the following, they arise when higher derivative (quartic) corrections to the gravity action are taken into account.  

Before considering the WYM case, as a warm-up we present the calculation of the topological susceptibility of finite temperature ${\cal N}=4$ SYM  at strong coupling. To our knowledge, this result was missing in the literature. Then we will move to the WYM model showing how a non-trivial instanton-driven temperature-dependent topological susceptibility arises.

To complete our analysis of the temperature dependence of the axion mass in WYM we will also study how much the axion coupling $f_a$ varies with $T$, referring to appendix \ref{app1} for details.

\subsection{Topological susceptibility of finite temperature ${\cal N}=4$ SYM}
Let us consider the ${\cal N}=4$ $\mathrm{SU}(N_c)$ SYM theory in $3+1$ dimensions. This is a superconformal gauge theory where the complex coupling $\tau = (2\pi)^{-1}\theta + 4\pi i g_\mathrm{YM}^{-2}$ is a modulus. There are no anomalies, hence, despite the fact that the theory contains massless (adjoint) fermions, the $\theta$-dependence cannot be rotated away. At zero temperature, however, the topological susceptibility of the theory is zero. This can be immediately deduced from dimensional analysis: the theory has no scale, while the susceptibility is dimensionful. The same result can also be obtained by a direct computation
\be
\chi_\mathrm{SYM}(T=0) \equiv \int d^4x \langle Q(x) Q(0)\rangle =0\,,
\ee
where $Q(x)\sim {\rm Tr} F\wedge F$ is the topological charge density operator, with protected conformal dimension $\Delta=4$ and the Euclidean correlator \footnote{It is worth recalling that the topological susceptibility is an equilibrium observable, not to be confused with the Chern-Simons diffusion rate, also known as Sphaleron decay rate, which is a transport coefficient related to a Wightman correlator of $Q(x)$. See {\it e.g.} \cite{moore} for a detailed discussion about this point.} is computed on the ground state at zero temperature. 

At finite temperature, however, the above result can well be modified. The holographic dual description is provided by a $AdS_5\times S^5$ black brane background
\be
ds^2 = \bfrac{r^2}{l^2}\left( - f(r) dt^2 + dx_i\,dx_i \right) + \bfrac{l^2}{r^2}\left(\bfrac{dr^2}{f(r)} + r^2 d\Omega_5^2\right)\,, \quad f(r) = 1-\frac{r_h^4}{r^4}\,,
\ee
with constant dilaton $e^{\phi}=g_s$ and a five-form flux on $S^5$ proportional to the number of colors $N_c	$. The radius of the horizon is related to the temperature by $r_h= \pi l^2 T$ and the $AdS_5$ radius is given by $l^4 = 4\pi g_s N_c \alpha'^2$.

The holographic picture suggests that the only possible contributions to the topological susceptibility can come from instanton 
corrections, {\it i.e.} $D$-instanton corrections on the gravity side \cite{massimo}. The reason is that the type IIB axion $C_0$, which is dual to the field theory $\theta$-angle according to the relation $\theta=2\pi C_0$, is constant on the background. To leading order in derivatives, the (related, truncated) type IIB effective action only contains derivatives of $C_0$, so that the on-shell gravity action will not show any $\theta$-dependent term. The situation changes if we include $D$-instanton corrections. As it was shown in \cite{gregu1} (see also \cite{gregu2}) these contribute to the first subleading correction (in $\alpha'$) to the type IIB effective action, {\it i.e.} to the so-called $R^4$ term. 
In Einstein frame and using the standard convention on the background value of the dilaton $e^{\phi}=g_s$, the latter can be written as (see {\it e.g.} \cite{gkt} and \cite{myersbuchel})
\be
\delta S = -\frac{1}{16\pi G_{10}}\int d^{10}x\sqrt{g}\,\alpha'^3 f(\tau,\bar \tau) g_s^{3/2}e^{-\frac32 \phi} W\,,
\label{deltar4}
\ee
where $W$ contains quartic terms in the Riemann tensor, 
 \be
\tau = C_0 + i e^{-\phi}\,,
\ee
and the non-holomorphic function $f(\tau,\bar \tau)$, in the $e^{\phi}\rightarrow0$ limit, is given by
\be
f(\tau,\bar \tau) = \frac{\zeta(3)}{8} + \frac{\pi^2}{24}e^{2\phi} + \frac{e^{2\phi}}{16} \sum_{N=1}^{\infty} G_{N,4}\,,
\label{fde}
\ee
where the first term arises at tree level, the second one at one loop and the third one contains the non-perturbative $D$-instanton corrections  (the summation runs over the $N$-instanton contributions). The leading order, one-instanton term reads \cite{gregu2}
\be
G_{1,4} = 4\pi e^{-\frac{\phi}{2}}e^{2\pi i \tau}+ c.c.
\label{G1}
\ee
The action term (\ref{deltar4}) has been computed at tree level in \cite{gkt}, to obtain the first subleading correction (in inverse powers of the 't Hooft coupling) to the free energy of ${\cal N}=4$ SYM at finite temperature in the holographic limit. As discussed in \cite{gkt} there exists a scheme where the quartic term $W$ can be written in terms of just the Weyl tensor 
\be
W=C^{hmnk}C_{pmnq}C^{rsp}_hC^{q}_{rsk}+\frac12C^{hkmn}C_{pqmn}C_h^{rsp}C^q_{rsk}\,.
\label{Wchoice}
\ee
With this choice, the term (\ref{deltar4}) does not modify the zero-temperature $AdS_5 \times S^5$ solution. However it does perturb the finite temperature black brane solution \cite{gkt}. Nevertheless, if one is interested in just the leading corrections to the field theory free energy, it is enough to compute the action term (\ref{deltar4}) on the unperturbed black brane solution. The related on-shell value of the quartic term $W$ has been computed in \cite{gkt} with the result
\be
W = \frac{180}{l^8} \frac{r_h^{16}}{r^{16}}\,.
\label{Wos}
\ee
Integrating over $S^5$ the on-shell action reduces to
\be
\delta S = -\frac{1}{16\pi G_5} \int d^5x \sqrt{g_5}\alpha'^3 f(\tau,\bar\tau) W \,,
\ee
where 
\be
16\pi G_5 = \frac{16\pi G_{10}}{\pi^3 l^5}\,,\quad \sqrt{g_5} = \frac{r^3}{l^3}\,.
\ee
Crucially, in view of the on-shell value of eq. (\ref{Wos}), one immediately realizes that the integral in $\delta S$ is perfectly convergent at large $r$ so that one does not need  to add any counterterm to the computation. 

Using the holographic relation $F = T S_{\rm{on-shell}}$ between the field theory free energy and the on-shell gravity action, we thus get that the correction to the free energy density corresponding to the $\delta S$ term reads\footnote{In this section we use different conventions w.r.t. those in \cite{gkt}
for what concerns the holographic definitions of the couplings. 
Here we use $\l_{\mathrm{YM}} = g_\mathrm{YM}^2 N_c = 4\pi g_s N_c$, while in \cite{gkt} $g_\mathrm{YM}^2 =2\pi g_s$.}
\be
\delta f = -\frac{15}{8} \pi^2 N_c^2 T^4 f(\tau,\bar\tau) \l_{\mathrm{YM}} ^{-3/2} \,.
\ee
The leading order $\theta$-dependent term arises from the one-instanton contribution to $f(\tau,\bar\tau)$ and reads (see eq.~(\ref{fde}) and (\ref{G1}))
\be
\delta f^{(\theta)} = - \frac{15}{128} \pi^{3/2} \sqrt{N_c}\, T^4 e^{-\frac{8\pi^2}{g_\mathrm{YM}^2}}\cos\theta\,,
\ee
so that the topological susceptibility is given by
\be
\chi_\mathrm{SYM}(T) = \frac{d^2 \delta f^{(\theta)}}{d\theta^2}|_{\theta=0} = \frac{15}{128} \, \pi^{3/2} \sqrt{N_c}\, T^4 e^{-\frac{8\pi^2}{g_\mathrm{YM}^2}}\,.
\ee
Notice that, apart from the overall factors, it has the same form as can be obtained for a dilute instanton gas. A crucial difference w.r.t.~to pure 
non supersymmetric Yang-Mills is that the gauge coupling does not run to leading order in $1/\l_{\mathrm{YM}}$ in the thermal case (actually the quartic term makes the dilaton running with $r$ \cite{gkt}). As a result, the topological susceptibility increases with $T$. 

Notice that the overall factor $\sqrt{N_c}$ is typical of instanton corrections in the present setup, see e.g.~\cite{Dorey:1999pd}.

\subsection{Topological susceptibility in the deconfined phase of WYM}
The WYM black brane solution (\ref{wymT}) in type IIA string theory can be obtained starting from the $AdS_7\times S^4$ black brane solution 
\be
ds^2= G_{MN}dx^{M}dx^{N} = \frac{y^2}{R^2} \left[ - f(y) dt^2 + \sum_{i=1}^4 dx_i^2 + d x_{10}^2 \right] + \frac{4R^2}{f (y) y^2} dy^2 + R^2 d \Omega_4^2,
\label{C1}
\ee
arising as the near horizon limit of the background sourced by $N_c$ non-extremal $M5$-branes. Here $f(y) = 1-y_0^6/y^6$ and the horizon position $y_0$ is related to the temperature by $3y_0=4\pi R^2 T$. The IIA background (\ref{wymT}) is obtained by reducing the above solution on a two-dimensional torus $S_{x_{10}}\times S_{x_{4}}$ with circles of radii $R_{4}= M_\mathrm{KK}^{-1}$ and $R_{10}= g_s l_s$ \cite{Witten:1998zw}. The solution also supports $N$ units of four-form flux along $S^4$. The flux quantization condition fixes $R^9/\kappa_{11}^2 = N_c^3/(2^7\pi^5)$, where $2\kappa^2_{11}=16\pi G_{N}^{(11)}=(2\pi)^8 l_{11}^9$ gives the 11-dimensional Newton constant. The 11-dimensional Planck length is related to the type IIA string scale by $l_{11}=g_s^{1/3} l_s$.

Quite remarkably, quartic corrections to the 11d supergravity action compactified on a torus are known, see e.g. \cite{greva,greguva}. With the conventions used in \cite{greguva,gkt} they read
\be
\delta S =  -\frac{1}{\kappa_{11}^{2/3}}\int d^{11}x \sqrt{-G}\,W\left[\frac{2\pi^2}{3}+{\cal V}_2^{-3/2} f(\rho,\bar\rho)\right]\,,
\label{r4M}
\ee
where $W$ can be expressed as in (\ref{Wchoice}), so that the extremal $AdS_7\times S^4$ solution is not modified by the action 
term (\ref{r4M}) \cite{gkt}. Moreover, (recalling that $R_4 = M_{\mathrm{KK}} ^{-1}$ and $g_{\mathrm{YM}}^2 = 4 \pi g_s l_s M_{\mathrm{KK}}$), 
\be
{\cal V}_2 = \frac{4\pi^2 R_{10}R_4}{\kappa_{11}^{4/9}} \sqrt{G_{(2)}} = \frac{4\pi^2 g_s l_s M_\mathrm{KK}^{-1}}{\kappa_{11}^{4/9}}\frac{y^2}{R^2} \equiv \frac{g_\mathrm{YM}^2}{M_\mathrm{KK}^2}\frac{\pi}{\kappa_{11}^{4/9}}\frac{y^2}{R^2} \,, 
\ee
is related to the volume $V_T$ of the torus by 
\be
V_T = \kappa_{11}^{4/9} {\cal V}_2 = \int dx_{4} dx_{10} \sqrt{G_{(2)}}\,,
\ee
with $G_{(2)}=G_{4\,4}G_{10\,10}$. In the ${\cal V}_2\rightarrow\infty$ limit the action term (\ref{r4M}) reduces to that considered in \cite{gkt} in the non-compact 11d case.

The modular function appearing in (\ref{r4M}) is defined as \cite{greva}
\be
f(\rho,\bar\rho) =  2\zeta(3) \rho_2^{3/2}+\frac{2\pi^2}{3}(\rho_2)^{-1/2}+ 4\pi (e^{2\pi i \rho}+e^{-2\pi i \bar\rho}) +\cdots\,,
\ee
where we neglect corrections with instanton number higher than one and we take the $\rho_2\rightarrow\infty$ limit with
\be
\rho\equiv \rho_1+ i \rho_2= (2\pi)^{-1}\theta + 4\pi i g_\mathrm{YM}^{-2}\,,
\ee
being proportional to the action (\ref{d0act}) of a Euclidean $D0$-brane wrapped along the $x_4$ circle \cite{greva}. 
The quartic term $W$ on the background (\ref{C1}) has been computed in \cite{gkt} and reads
\be
W= \frac{3285}{64 R^8}\frac{y_0^{24}}{y^{24}}\,.
\ee
Using the above expressions we get that the free energy density of the WYM theory at $T>T_c$ receives the following contributions from the quartic term (\ref{r4M})
\be
\delta f = \delta f_\mathrm{GKT} + \delta f_{{\cal V}_2}\,,
\ee
where
\be
\delta f_\mathrm{GKT} = - 730 \pr{ \frac{2 \pi}{3} } ^8 \pr{\frac{\pi}{2}}^{4/3}  \lambda_{\rm{eff}}(T)T^4\,,
\ee
can be obtained by a simple compactification of the related M5-brane result found in \cite{gkt} and
\be
\lambda_{\rm{eff}}(T) \equiv g_\mathrm{YM}^2 N_c \frac{T^2}{M_\mathrm{KK}^2} \equiv \l_{\mathrm{YM}} \frac{T^2}{M_\mathrm{KK}^2}\,.
\ee
The novel contribution is
\be
\delta f_{{\cal V}_2} = - \frac{3285}{42} \pr{\frac{2 \pi}{3}}^4  M_\mathrm{KK}T^3\left[\frac{32\pi^2 N_c^2}{\l_{\mathrm{YM}}^2}\zeta(3) +\frac{2\pi^2}{3} + 16\pi^{3/2}\frac{\sqrt{N_c}}{\sqrt{\l_{\mathrm{YM}}}} e^{-8\pi^2/g_\mathrm{YM}^2}\cos\theta\right]\,.
\ee
Hence, from the $\theta$-dependent term, we see that,  to leading order in the instanton expansion, the topological susceptibility of the WYM model in the deconfined phase reads
\be\label{suscwym}
\chi_\mathrm{WYM}(T) =  \frac{3285 \pi^{3/2}}{42}  \bigg(\frac{4 \pi}{3}\bigg)^{\!4} \frac{\sqrt{N_c}}{\sqrt{\lambda_{\rm{eff}}(T)}}\; T^4 e^{-\frac{8\pi^2}{g_\mathrm{YM}^2}}\,.
\ee
Notice that we get the same overall $\sqrt{N_c}$ factor as in the SYM case and that the scaling with the temperature is given by $\chi_\mathrm{WYM}\sim M_\mathrm{KK}T^3$. 

Let us conclude, for completeness, by recalling the status of the analogous computation in the alternative background, dual to the deconfined phase, presented in \cite{Mandal:2011ws}.
In \cite{Hanada:2015gsa} it is pointed out that extracting the instanton action for generic instanton size is not possible at present.
So, a complete estimate of the topological susceptibility for that background is lacking.
What can be done is to check that, in this case, the dual instanton action  is peaked at a specific size, where the temperature behavior is of the form $S_{inst} \sim \frac{8 \pi^2}{g_\mathrm{YM} ^2} (1 - {\rm const}\sqrt{T_c/T})$.

\subsection{Temperature dependence of the axion mass in WYM}\label{sec:Tdepmass}

The temperature dependence of the axion mass in the deconfined phase of WYM can be read from a suitable generalization of eq.~(\ref{mWVaxion})
\be
m_{a}^2(T) = \frac{2}{f_{a}^2(T)}\chi_\mathrm{WYM}(T)\,.
\label{mWVaxionT}
\ee	
If the temperature is much smaller than $f_a$ (the zero-temperature axion coupling) but higher than the deconfinement temperature $T_c$, the temperature dependence of the axion coupling is usually neglected. 
In the WYM model the coupling $f_a(T)$ can be deduced by a careful extension to the deconfined WYM background of the computations reviewed in appendix \ref{app1} for deducing $f_a$. The details are provided at the end of that appendix. The result is that, in the $T_c<T\ll f_a$ regime, {\it i.e.} for $LT_c<L T\ll1$
\begin{equation}\label{faTqui}
f_a^2(T)\approx f_a^2 \left[1+\frac{1.3}{\pi^6} \lambda^2 N_c^2 \frac{T^6}{M_{KK}^2 f_a^4}+\cdots \right]\,,
\end{equation}
where we have used eq. (\ref{faT0}) for the zero temperature axion coupling $f_a$.
So the decay constant $f_a(T)$ slightly increases with temperature in that regime. Just as for the topological susceptibility, this behavior differs from what is expected in QCD. At very high temperatures, instead, (more precisely at $L T\ge L T_{\chi}\approx0.154$ \cite{Aharony:2006da}) the axion melts as a result of the fact that the energetically favoured configuration for the PQ D8-brane corresponds to two disconnected branches.

Thus, in the $LT\ll1$ regime, the temperature dependence of $f_a$ can be neglected and that of the axion mass in the deconfined phase of the WYM model is driven by that of the topological susceptibility (\ref{suscwym}). The result is that $m_a^2$ increases like $T^3$ in that regime. This behavior is of course very different from the power-like suppression of the axion mass with temperature which can be extracted, from asymptotic freedom, in Yang-Mills in the dilute instanton gas approximation. The peculiar higher dimensional UV completion of the WYM model is at the basis of this expected discrepancy.


\section{Conclusions} \label{seccomments}

In this paper we have introduced a new calculable, strongly coupled UV completion of the low energy QCD axion physics in the planar limit. 
The completion is provided by the five dimensional theory (and then by the six dimensional (2,0) theory in the M-theory limit) in the UV of the holographic WSS model.
The construction seems to evade the phenomenologically unsatisfactory constraints on the allowed values of $f_a$ commonly encountered in string theory axion models \cite{wittenaxion}.

Often a higher dimensional embedding of the axion physics provides a natural protection against higher dimensional operators which could spoil the PQ mechanism (see e.g.~\cite{acharya}). It would be interesting, for the future, to check if this is the case in the present model.

Since in the construction the Peccei-Quinn symmetry is unrelated to Standard Model matter, the model falls in the KSVZ class.
The couplings of the axion with the nucleons and the photon have no UV contributions and are entirely determined by the low energy action and the mixing with the pseudoscalars.
The magnitude of the couplings to nucleons have been estimated in \cite{axioncouplings}.

In the deconfined phase, we have evaluated the topological susceptibility, the temperature dependence of $f_a$ and that of the axion mass. The latter is found to be an increasing function of the temperature.
The setting seems to be less reliable as a QCD model in this phase, for it exhibits a higher dimensional completion and absence of asymptotic freedom.

As an aside, we have calculated for the first time the topological susceptibility of ${\cal N}=4$ SYM at strong coupling.

A particularly interesting task for the future would be the study of axionic strings and domain walls, since the holographic model geometrizes these topological objects \cite{dubovsky}. It would be also interesting to investigate the consequences of the presence of more than one extra flavor in the axion sector.

\vskip 15pt \centerline{\bf Acknowledgments} \vskip 10pt \noindent 
We are grateful to Massimo Bianchi for illuminating discussions on instanton corrections in string theory and for observations. We thank Riccardo Argurio, Michele Redi, Andrea Tesi, Pierre Vanhove for comments and discussions.  We thank the Galileo Galileo Institute (Florence) and the organizers of the workshop ``String Theory from a Worldsheet Perspective'' during which part of this work has been completed. This work was supported by the Simons Foundation grant 4036 341344 AL. 
\appendix
\section{The WSS model with antipodal $D8$-branes}
\label{appreview}
\setcounter{equation}{0} 

Let us consider adding $N_f$ probe $D8$-branes to the WYM background (\ref{wym}). The induced metric on their worldvolume can be conveniently written in terms of the coordinates \cite{Sakai:2004cn}
\begin{equation}
u^3 = u_{0}^3 + u_{0} {\tilde u}^2\;,\quad \varphi = \frac{2\pi}{\beta_4} x_4 \,,\label{defUcoord}
\end{equation}
and, in turn, parameterizing the $({\tilde u},\varphi)$ plane by coordinates $(y,z)$
\begin{equation}
y=\tilde u\cos\varphi\;,\quad z=\tilde u \sin \varphi\,.
\label{yzdef}
\end{equation}
In this way the cigar metric reads
\begin{equation}
d s^2_{(y,z)} = \frac{4}{9}\left(\frac{R}{u}\right)^{3/2}\left[\left(1-q(\tilde u)z^2\right)d z^2 + \left(1-q(\tilde u)y^2\right)d y^2 - 2zy \, q(\tilde u)d z d y\right]\,,
\end{equation}
with $u = u(z,y)$ and $q(\tilde u)= \frac{1}{{\tilde u}^2}\left(1-\frac{u_{0}}{u}\right)$. 
The antipodal $D8$ embedding is simply given by $y=0$. The stability of the configuration with respect to small deformations of the embedding has been demostrated in \cite{Sakai:2004cn}. Note that $z\in (-\infty,\infty)$ and that the two asymptotic values correspond to the two $D8-\overline{D8}$ branches at antipodal points on the $x_4$ circle. The $D8$-brane embedding, therefore, nicely realizes the chiral symmetry breaking $\mathrm{U}(N_f)\times \mathrm{U}(N_f)\rightarrow \mathrm{U}(N_f)$ of the dual field theory.

The $D8$-branes action, at leading order in the low energy expansion reads
\begin{equation}\label{SYM}
S_\mathrm{WSS} = -\kappa\int d^4x d z\, \left(\frac{1}{2}h(z)  \, \Tr\mathcal{F}_{\mu\nu}\mathcal{F}^{\mu\nu} +   k(z)  \Tr\mathcal{F}_{\mu z}\mathcal{F}^\mu_{\;\; z}\right)
 + \frac{N_c}{24\pi^2}\int  \omega_5(\mathcal{A}) \,,
\end{equation}
where (in units $u_0=1$)
\be
\kappa = \frac{N_c\lambda}{216\pi^3}M_\mathrm{KK}^2
\,,\quad h(z) = (1+z^2)^{-1/3}\,,\quad k(z) = (1+z^2)\,,
\ee
and 
\begin{equation}
\omega_5(\mathcal{A}) =  \Tr\left(\mathcal{A}\wedge \mathcal{F}^{ 2}- \frac{i}{2}\mathcal{A}^{ 3} \wedge\mathcal{F} - \frac{1}{10}\mathcal{A}^{ 5}\right)\;,\quad d \omega_5(\mathcal{A}) = \Tr\mathcal{F}^{  3}\,.
\label{omega5}
\end{equation}
Here ${\cal A}$ is the $\mathrm{U}(N_f)$ gauge field on the $D8$-branes. As we have recalled in section (\ref{secaxion}), the fluctuations of this gauge field on the background are holographically related to mesonic fields in the dual field theory. Instanton solutions correspond instead to baryons.

Let us restrict for a moment to the Abelian case $N_f=1$, and insert into (\ref{SYM}) the following expansions for the gauge field 
\bea\label{expA}
&&{\mathcal A}_{z} (x^{\mu},z) = \sum_{n=0}^{\infty} \varphi^{(n)}(x^{\mu})\phi_n(z)\,,\nonumber \\ 
&&{\mathcal A}_{\mu}(x^{\mu},z) = \sum_{n=1}^{\infty} B_{\mu}^{(n)}(x^{\mu})\psi_n(z)\,.
\eea
Choosing the functions $\phi_n(z)$, $\psi_n(z)$ to form complete, suitably normalized sets
\be \label{eqforpsi}
-h(z)^{-1} \partial_z (k(z)\partial_z\psi_n(z))=\lambda_n \psi_n(z)\,, \qquad \kappa  \int dz \,h(z) \psi_n(z) \psi_m(z) = \delta_{mn}\,,
\ee
and inserting (\ref{expA}) in (\ref{SYM}) gives the following four dimensional action \cite{Sakai:2004cn}
\begin{equation}
\begin{aligned}
S &=-\kappa\int d^4x \, \left[\sum_{n\geq 1}\left(\frac{1}{4}  \mathcal{F}^{(n)}_{\mu\nu}\mathcal{F}^{\mu\nu(n)}+  \frac{1}{2} m_n^2 B_\mu^{(n)}B^{\mu(n)}\right)+  \frac{1}{2}\partial_\mu \varphi^{(0)} \partial^\mu \varphi^{(0)}\right]\,.
\end{aligned} \label{kineticmesons}
\end{equation}
Under parity transformation $P:(x^\m,z) \rightarrow (-x^\m,-z)$ we have $(\mathcal{A}_\m,, \mathcal{A}_z ) \rightarrow (- \mathcal{A}_\m,, - \mathcal{A}_z )$, 
so from (\ref{expA}) we see that $B^{(n)} _\m$ are four-dimensional vectors (resp.~axial vectors) when $\psi_n$ is even (resp.~odd) while $\f^{(n)}$ are 
four-dimensional scalars (resp. pseudoscalars) 
when $\ff_n$ is odd (resp. even).
The $B_{\mu}^{(n)}$ modes correspond to massive vectors (resp.~axial vectors), for odd (resp.~even) $n$, with masses $m_n^2 = \lambda_n M_\mathrm{KK}^2$. For 
example, $B_{\mu}^{(1)}$ is identified with the $\rho$ meson and $B_{\mu}^{(2)}$ with the $a_1$ meson. The scalar modes $\varphi^{(n)}$ for $n\ge1$ get eaten 
by the $B_{\mu}^{(n)}$, and the massless field $\varphi^{(0)}$ is associated to the mode $\psi_0 \propto (1+z^2)^{-1}$ which is an even function: it is thus a 
pseudoscalar field, the $N_f=1$ version of the Goldstone bosons arising from the spontaneous breaking of chiral symmetry. Other massive scalar mesons are 
given by fluctuations of the $D8$-brane embedding. 

Thus, the effective action (\ref{SYM}) includes into a unified picture both the low lying mesonic modes and the whole tower of massive mesons. Their properties (masses, decay rates, couplings) are given in terms of the few parameters of the model, {\it i.e.} $N_c,N_f,M_\mathrm{KK}$ and $\lambda$. Clearly, as in QCD, the low energy physics is dominated by the pions, since all the other modes (glueballs, Kaluza-Klein modes and axial or vector mesons) have masses of order $M_\mathrm{KK}$.

\subsection{Chiral Lagrangian}
Suitably generalizing the discussion above to the $N_f>1$ case, the low energy effective action on the $D8$-branes reduces to \cite{Sakai:2004cn}
\begin{equation}\label{chiL}
S =  \kappa \int d^4x\,\Tr\left(a\,(U^{-1} \partial_\mu U)^2 + b\, ([U^{-1} \partial_\mu U,U^{-1} \partial_\nu U])^2\right)\,,
\end{equation}
where $a$ and $b$ are constants given by
\begin{equation}
a = \frac{1}{\pi}\;,\quad b  = \frac{1}{2\pi^4}\cdot 15.25\ldots\,,
\end{equation}
and\footnote{Here we have used the gauge ${\cal A}_z=0$. We have also ignored the (massive) vector mesons in the expansion of 
${\cal A}_\mu= i U^{-1} \partial_\mu U \,\psi_+ + \sum_{n} B_\mu^{(n)}(x_\mu)\psi_n(z)$. 
The vector mesons can be consistently added to the Chiral Lagrangian \cite{Sakai:2004cn}.}
\begin{equation}
\mathcal{A}_\mu(x^\mu,z) = i U^{-1} \partial_\mu U \,\psi_+\,,
\end{equation}
with $\psi_\pm(z) = \frac{1}{2}(1\pm \psi_0(z))$ and
\begin{equation}
U(x^\mu) = \mathcal{P} \exp\left( i \int_{-\infty}^\infty d z\, \mathcal{A}_z(x^\mu,z)\right)=\exp \left(\frac{2i}{f_\pi}\pi^a(x^\mu) T^a\right)\,,
\label{pionm}
\end{equation}
being the pseudoscalar meson matrix. Here $T^a$ are $\mathrm{U}(N_f)$ generators normalized to $\Tr(T^aT^b)=\frac{1}{2}\delta_{ab}$.

The action (\ref{chiL}) precisely coincides with the chiral Lagrangian including the Skyrme term. The pion decay constant $f_{\pi}$ and the Skyrme coupling $e$ are given by
\begin{equation}
f_{\pi} = 2 \sqrt{\frac{\kappa}{\pi}}\,, \qquad e^2 \sim \frac{1}{2.5\kappa}\,.
\label{fpaie}
\end{equation}
\subsection{Theta and mass terms}
As we have recalled in section \ref{secaxion}, turning on a topological $\theta$-term in the unflavored gauge theory amounts to turning on a non trivial $F_2=dC_1\sim\star dC_7$ RR two form with non-zero flux along the cigar in the WYM background (\ref{wym}). The backreaction of $F_2$ can be neglected if $\theta/N_c\ll1$. If we further add to the background $N_f$ $D8$-brane probes, the equations of motion and the Bianchi identity for the RR form get modified, due to the Chern-Simons coupling $\int C_7\wedge \Tr{\cal F}$ on the $D8$-brane worldvolume. This is accounted for by the bulk+branes action term
 \begin{equation}
S_{C_7} = -\frac{1}{4\pi}(2\pi l_s)^{6}\int d C_{7}\wedge \star d C_{7} + \frac{1}{2\pi}\int C_{7} \wedge \Tr{\cal F}\wedge \omega_y\,,
\label{sc7}
\end{equation}
where we have introduced the one-form $\omega_y = \delta(y)d y$, in order to extend the $D8$ integral to the whole spacetime and we have used the same normalization on the RR fields as in \cite{Sakai:2004cn}. The equation of motion for $C_7$ can be rewritten as the modified Bianchi identity for $\tilde F_2 = (2\pi l_s)^6 dC_7$
\be
d \tilde F_2 = \Tr \mathcal{F} \wedge\delta(y) d y\,.
\label{bianchi}
\ee
This can be formally solved by
\be
\tilde F_2 = dC_1 + \sqrt{\frac{N_f}{2}}\widehat{A} \wedge\delta(y) d y\,,
\label{intB}
\ee
where $\widehat{A}$ is the Abelian component of the $\mathrm{U}(N_f)$ gauge field, {\it i.e.} the hatted field in the decomposition
\begin{equation}
\mathcal{A} = \widehat{A} \frac{1}{\sqrt{2N_f}} + A^a T^a\,,
\label{separ}
\end{equation}
where $T^a$ are now the $\mathrm{SU}(N_f)$ generators.

The action $S_{C_7}$ is actually equivalent to \cite{Bartolini:2016dbk}
\be
S_{{\tilde F}_2} = -\frac{1}{4\pi (2\pi l_s)^6} \int d^{10}x |\tilde F_2|^2\,.
\ee
Considering a zero mode for ${\widehat A}_z$ such that
\be
\int dz {\widehat A}_z = \frac{2 \etaprime}{f_{\pi}}\,, 
\ee
using the integrated Bianchi identity for $\tilde F_2$ (on the cigar directions) and the equation of motion $d\star \tilde F_2 = 0$, the on-shell value of the action above reduces to
\be\label{SF2}
S_{{\tilde F}_2} = -\frac{\chi_\mathrm{WYM}}{2}\int d^4x\left(\theta + \frac{\sqrt{2N_f}}{f_{\pi}}\etaprime\right)^2\,,
\ee
where $\chi_\mathrm{WYM}$ is the topological susceptibility of the unflavored model (\ref{paraWSS}). From this action one can also deduce the large $N_c$ estimate of the $\etaprime$ mass predicted by the Witten-Veneziano formula
\be
m_\etaprime ^2 = m_\mathrm{WV}^2\equiv\frac{2N_f}{f_{\pi}^2}\chi_\mathrm{WYM}\,.
\ee
The $\theta$-dependence in (\ref{SF2}) can actually be rotated away by a chiral rotation, which corresponds to a gauge transformation of the Abelian field $\widehat A$ \cite{Sakai:2004cn}
\begin{equation}\label{shift}
\frac{1}{\sqrt{2N_f}}\int \widehat{A}_z \;\longrightarrow\; \frac{1}{\sqrt{2N_f}}\int \widehat{A}_z -\frac{\theta}{N_f}\,.
\end{equation}
This is not unexpected since the quarks we are adding by means of the WSS $D8$-branes are massless. 

Actually, in the WSS model it is possible to turn on a mass term for the quarks of the same form $\Tr [M U^\dagger + h.c.]$ as in the chiral Lagrangian setup. It reads
\begin{equation}\label{Smass}
S_\mathrm{mass} = c \int d ^4x\, \Tr  \mathcal{P} \left[M  \exp\left(-i\int_{-\infty}^\infty\mathcal{A}_z d z\right) + h.c. \right]\,,  
\end{equation}
where $c$ is a constant and $M$ is the flavor mass matrix. This term has actually a very precise meaning in string theory 
\cite{AK, Hashimotomass}: it is the deformation due to open string worldsheet instantons stretching between the $D8$-branes. The deformation 
has the simple form given above only for quark masses parametrically much smaller than $M_\mathrm{KK}$. Moreover it is 
trivially zero in the deconfined phase where chiral symmetry is restored. It is easy to realize that adding the action above to 
the action (\ref{chiL}) gives masses to the pions according to the GMOR relation (\ref{GMOR}) \cite{Bartolini:2016dbk}.

Now, after the shift (\ref{shift}), $S_\mathrm{mass}$ becomes
\begin{equation}
S_\mathrm{mass} = c \int d ^4x\, \Tr \mathcal{P} \left[M e^{i\frac{\theta}{N_f}}  \exp\left(-i\int_{-\infty}^\infty\mathcal{A}_z d z\right) + h.c.\right]\,, 
\label{smass}
\end{equation}
which amounts to redefining
\begin{equation}
M  \;\longrightarrow \; M e^{i\theta/N_f}\,.
\end{equation}
The $\theta$-dependence is thus not erased anymore. Moreover, as expected in QCD, the physical $\theta$ parameter is given by the combination
\begin{equation}
\bar\theta = \theta + \arg \det M\,.
\label{thetaphys}
\end{equation}

Combining together the WSS effective action (\ref{chiL}), the action (\ref{SF2}) and the mass term (\ref{smass}) and expanding to lowest order in derivatives, one gets precisely the effective Lagrangian (\ref{leffe}) without the axion term ($a=0$). 

As we have discussed in section \ref{secaxion}, the axion field is introduced in the model by means of an extra $D8$-brane (which we have called Peccei-Quinn, PQ, $D8$-brane) which is not antipodally embedded. The following section is thus devoted to a review of non-antipodally embedded $D8$-branes in the model.

\section{Non-antipodally embedded $D8$-branes}\label{app1}
\setcounter{equation}{0}
Let us consider embedding in the WYM background (\ref{wym}) a PQ $D8$-brane with two asymptotic branches placed at non-antipodal points on the $x_4$ circle. Their distance on the circle is thus $L < \pi R_{4}$, {\it i.e.} $L M_\mathrm{KK} < \pi$. 
The part of the DBI action which depends on the metric reads
\ba
S_{D8} = -\t_8 V_{3+1} V_4 \int dx_4 u^4 \sqrt{f(u)+\pr{\frac{R}{u}}^{3} \frac{u'^2}{f(u)}} \ ,
\ea
where 
\be
\t_p = \frac{1}{(2\pi)^{p} l_s^{p+1}} \ , \q \q u' = \frac{du}{dx_4}\,,
\ee
and $V_{3+1}$ and $V_4$ are, respectively, the field theory spacetime volume and the $S^4$ volume.
 Since the 
Lagrangian does not explicitly depend on $x_4$ we have a first integral:
\be
\label{firstintegral}
\frac{u^4 \sqrt{f(u)}}{\sqrt{1+\pr{\frac{R}{u}}^3 \frac{u'^2}{f^2 (u)}}} = u_J	 ^4 \sqrt{f(u_J)} \ .
\ee
Here $u_J$ is the point where $u'$ vanishes, as illustrated in figure \ref{ssa}.
The distance $L$ is given by
\ba
L &=& \int d x_4 = 2 \int_{u_J} ^\infty \frac{d u}{u'} = 2 \int _{u_J} ^\infty \frac{du}{f(u) \pr{\frac{u}{R}}^{3/2} 
\sqrt{\pr{\frac{u}{u_J}}^8 \frac{f(u)}{f(u_J)}-1}} \nb \\
&=& \frac{2}{3} R^{3/2} u_J ^{-1/2} \sqrt{1-b^3} \int_0 ^1 dy \frac{y^{1/2}}{\big(1-b^3 y\big)\sqrt{1-b^3 y -(1-b^3)y^{8/3}}} \ .	\label{Lint}
\ea
In the last line we substituted $y= u_J^3/u^3$ and defined $b = u_0/u_J$.
Calling
\be
\label{defintegralj}
J(b) = \frac{2}{3} \sqrt{1-b^3} \int_0 ^1 dy \frac{y^{1/2}}{\big(1-b^3 y\big)\sqrt{1-b^3 y -(1-b^3)y^{8/3}}}  \ , 
\ee
we can write
\be
\label{elleinfunzionediuzero}
L = J(b) R^{\frac32} u_J ^{-\frac12} \ .
\ee

Let us now show that the Abelian gauge theory on the PQ-brane gives a pseudoscalar pseudo-Nambu-Goldstone boson (which we will identify with the axion field $a$) analogous to the mode $\varphi^{(0)}$ in the WSS $N_f=1$ antipodal case, see eq.~(\ref{kineticmesons}). On top of this there will also be a tower of massive vector mesons. As we will show, the mass of the latter scales as $m^2 \sim L^{-2}$ and therefore 
the vector mesons decouple at low energy \cite{Aharony:2006da}. 

Expanding the $D8$-brane DBI action we obtain the term
\be
S_{\mathcal{FF}} = -\frac{\t_8}{4} (2 \pi \a')^2 \int d^9 x e^{-\ff} \sqrt{\det g_{MN}} g^{MR} g^{NS} \mathcal{F}_{MN} \mathcal{F}_{RS} \ .
\ee
Integrating over $S^4$ with the assumption that the field strength $\mathcal{F} _{MN}$ does not depend on the $S^4$ coordinate we get
\ba
\label{axionactionbeforemodes}
S_{\mathcal{FF}} =-C \int d^4 x du 
 \g^{-1/2} \Big( R^3 u^{-1/2} \mathcal{F}_{\m \n} \mathcal{F}^{\m \n}+
 2 \g u^{5/2}  \eta^{\m \n} \mathcal{F}_{\m u} \mathcal{F}_{\n u}\Big)  \ ,
\ea
where we defined
\ba
C &=& \frac{\t_8 V_{4} R^{3/2}}{2 g_s} (2 \pi \a')^2
\ , 	\label{constC}\\
\g (u) &=& f(u)-f(u_J) \pr{\frac{u_J}{u}}^8 \ .
\ea
It is convenient to work with the coordinate $z$, defined by
\be
u = \pr{u_J ^3 + u_J z^2}^{1/3} \ .
\ee
This coordinate takes values along the whole real axis, it is positive for $x_4 > L/2$ and negative for $x_4 < L/2$.
The action then reads 
\ba
\label{axionactionbeforemodeszcoordinate}
S_{\mathcal{FF}} &=&-\frac{C u_J}{3} \int d^4 x dz 
 \Big( \frac{R^3 \g^{-1/2}|z|}{u^{5/2}}   \mathcal{F}_{\m \n} \mathcal{F}^{\m \n}+
 \frac{9}{2} \g^{1/2} \frac{u^{9/2}}{u_J ^2 |z|}  \eta^{\m \n} \mathcal{F}_{\m z} \mathcal{F}_{\n z}\Big)  \ .
\ea
Following \cite{Sakai:2004cn} we expand the components of the gauge field as
\begin{subequations}
\ba
A_\m (x^\m, z) &=& \sum_n B_\m ^{(n)} (x^\m) \psi _n (z) \ , \\
A_z (x^\m, z) &=& \sum_n \f ^{(n)} (x^\m) \ff _n (z) \ .
\label{componentAz}
\ea
\end{subequations}
Accordingly,
\begin{subequations}
\ba
\mathcal{F}_{\m \n} (x^\m, z) &=& \sum_n F_{\m \n} ^{(n)}(x^\m) \psi _n (z) \ , \\
\mathcal{F}_{\m z} (x^\m, z) &=&\sum_n \p_\m \f ^{(n)} (x^\m) \ff _n (z) - \sum_n B_\m ^{(n)} (x^\m) \p_z \psi _n (u) \ . \q \q
\ea
\end{subequations}

\subsection{PQ vector mesons}
Let us first consider only the vector mesons. We have 
\ba
S_{\mathcal{FF}} ^{\text{vec}} &=&-\frac{C u_J}{3} \sum_{n,m} \int d^4 x dz  
  \Big(  \frac{R^3 \g^{-1/2}|z|}{u^{5/2}}   F^{(n)} _{\m \n} F^{(n)\m \n} \psi _m \psi_n +\nb \\
 &+&
 \frac{9}{2} \g^{1/2} \frac{u^{9/2}}{u_J ^2 |z|}   B_\m ^{(n)} B^{(m) \m} \p_z \psi_n \p_z \psi_m \Big)  \ .
\ea
In order to obtain the kinetic term of the massive vector action we ask $\psi_n$ to be normalized such that
\be
\frac{4 C u_J R^3}{3} \int  _{-\infty} ^{+\infty} dz |z| \g^{-1/2} u^{-5/2} \psi _n \psi_m = \d _{m n} \ .
\ee
Let us take the mass term
\ba
&&- \frac{3C}{2 u_J} \sum_{n,m} \int dz \frac{\g^{1/2}}{|z|}  u^{9/2}  \p_z \psi_n \p_z \psi_m  B_\m ^{(n)} B^{(m) \m} = \nb \\
&=& \frac{3C}{2 u_J}  \sum_{n,m} \int dz \p_z \pr{ \frac{\g^{1/2}}{|z|} u^{9/2}  \p_z \psi_n  } \psi_m  B_\m ^{(n)} B^{(m) \m} \ .
\ea
In order to obtain the mass term for the tower of mesons we ask
\be
\label{nonantipodalvectormassterm}
-\frac{9}{4 u_J ^2} \frac{\g^{1/2} u^{5/2}}{|z|}
\p_z \pr{ \frac{\g^{1/2}}{|z|} u^{9/2}  \p_z \psi_n } = m_n ^2 R^3 \psi_n \ ,
\ee
so that
\ba
\label{vectormasstermnormalizationzcoordinate}
\frac{3C}{2 u_J} \int dz \frac{\g^{1/2}}{|z|}  u^{9/2}  \p_z \psi_n \p_z \psi_m   = \frac{1}{2} m_n ^2 \d _{m,n} \ .
\ea
By dimensional analysis, in the limit $u_J \gg u_0$ we have
\be
m_n ^2 \sim \frac{u_J}{R^3} \sim \frac{1}{L^2} \ .
\ee
\subsection{Scalar sector: the holographic axion}
From the second term in (\ref{axionactionbeforemodeszcoordinate}), 
\ba
S_{\mathcal{FF}} ^{\text{scal}} =-\frac{3C}{2 u_J} \int d^4 x dz   
  \g^{1/2} \frac{u^{9/2}}{|z|}  \eta^{\m \n} \mathcal{F}_{\m z} \mathcal{F}_{\n z}  \ , 
\ea
we read the kinetic term for the tower of scalar mesons:
\be
S_{\mathcal{FF}} ^{\text{scal}} = -\frac{3C}{2 u_J} \sum_{m,n} \int d^4 x dz   
  \g^{1/2} \frac{u^{9/2}}{|z|}  \p_\m \f^{(m)} \p^\m \f^{(n)} \ff _m \ff_n \ .
\ee
In order to canonically normalize the fields $\f ^{(m)}$ we ask that $\ff _m$ satisfy the orthonormality condition
\be
\label{scalarkinetictermnormalizationzcoordinate}
\frac{3C}{ u_J} \int  dz   
  \g^{1/2} \frac{u^{9/2}}{|z|}  \ff _m \ff_n = \d_{m,n} \ .
\ee
Comparing (\ref{vectormasstermnormalizationzcoordinate}) and (\ref{scalarkinetictermnormalizationzcoordinate}), we see that it is possible to choose 
$\ff_n = m_n ^{-1} \p_z \psi_n$.
There exists a zero mode,
\be
\label{zeromodeaxion}
\ff_0 (z) = C_0 |z| \g^{-1/2} u^{-9/2} \ , \q \q C_0 = \pr{\frac{3C}{u_J} \int _{-\infty} ^{+\infty} dz |z| \g^{-1/2} u^{-9/2} }^{-1/2} \ ,
\ee
which is orthogonal to the other modes, since
\be
\int dz \g^{1/2} \frac{u^{9/2}}{|z|}  \ff _m \ff_0 \sim \int dz \p_z \psi = 0 \ .
\ee
The field strength $\mathcal{F}_{\m z}$ is then decomposed as
\be
\mathcal{F}_{\m z} = \p_\m \f ^{(0)} \ff_0 + \sum _{n \geq 1} \pr{m_n^{-1}\p_\m \f ^{(n)} - B_{\m} ^{(n)}} \psi_n  \ .
\ee
We can choose a gauge in which the second term vanishes, hence we are left with the action of a spinless massless field,
\be
\label{axkin}
S_{\mathcal{FF}} ^{\text{scal}} = -\frac{1}{2}\int d^4 x \p_\m \f ^{(0)} \p^\m \f ^{(0)} \ .
\ee
Under parity transformation $P:(x^\m,z) \rightarrow (-x^\m,-z)$ we have $A_z \rightarrow -A_z$, so, since $\ff_0 (z)$ is an even 
function, from (\ref{componentAz}) we see that $\f^{(0)}$ must be a pseudoscalar field. We are thus led to identify 
$\f^{(0)}$ with the axion field $a$. As we will see, the axion field gets a Witten-Veneziano mass precisely as the 
$\etaprime$ in the previous section.

To summarize, the gauge field on the PQ $D8$-brane gives rise to the axion field and to a tower of PQ vector mesons,
\be
S_{\mathcal{FF}} = -\int d^4 x \pq{\frac{1}{2} \p_\m a \p^\m a +\sum_{n \geq 1} \pr{ 
\frac{1}{4} F_{\m \n} F^{\m \n} + \frac{1}{2} m_u ^2 B_\m ^{(n)}  B^{(n)\m}
}} \ .
\ee

\subsubsection{The axion decay constant $f_a$}
In order to find the axion decay constant $f_a$, namely the analogous of $f_\pi$ for the axion chiral Lagrangian, we define 
the holonomy for the axion field as
\be	\label{holonomyV}
V(x^\mu) = \mathcal{P} \exp \pr{ i \int_{-\infty}^\infty d z\, A_z(x^\mu,z) }
=\exp \pr{ \sqrt{2}i \frac{\f ^{(0)}}{f_a} }  \ .
\ee
Expanding these exponentials up to the linear term and using the expression (\ref{componentAz}) for $A_z$ and 
(\ref{zeromodeaxion}) for the zero mode we have
\be	\label{V1storder}
V \sim 1 + i \int _{-\infty} ^{+\infty} dz A_z =1+ i C_0 \f^{(0)} \int _{-\infty} ^{+\infty} dz |z| \g^{-1/2} u^{-9/2}
\ee
and therefore 
\be
f_a ^2 =   \frac{3C}{u_J} \pr{\int_{0} ^\infty dz z \g^{-1/2} u^{-9/2}}^{-1}  \ .
\ee
Factorizing the dimensionful quantities by the change of variable~$y= u_J^3/u^3$, we obtain
\be
f_a ^2 = \frac{6C}{ I(b)} u_J ^{3/2}  \ , 
\ee
where
\be
I(b) = \int_0 ^1\!dy\ \frac{y^{-\frac12}}{\sqrt{1-b^3y -(1-b^3)y^{\frac83}\,}} \ .
\ee
The function $I(b)$ is of order one for any value of $b<1$.
Finally, in terms of field theory quantities the axion decay constant reads
\be
f_a ^2 = \frac{N_c \l}{16 \pi^3 } \frac{J^3(b)}{I(b)}  \frac{1}{M_\mathrm{KK} L^3}  \ .
\label{fares}
\ee
Notice that for $b\ll1$
\be
f_a^2 \approx 0.1534 \frac{N_c \l}{16 \pi^3} \frac{1}{M_\mathrm{KK} L^3} +{\cal O}(b^3) \,.\label{fa0bsmall}
\ee
As we have seen in section \ref{secaxion}, phenomenological constraints actually require $b$ to be tiny. Hence, the leading order term in (\ref{fa0bsmall}) can be taken as the defining relation between the zero temperature axion coupling and the parameters of the WSS model. Using the Witten-Veneziano formula for the axion mass we can verify that in the regime in which we work the vector bosons masses are much bigger than the axion one:
\ba
\frac{m_a ^2}{m_n ^2} \sim \frac{ \chi_\mathrm{WYM} L^2}{f_a ^2} \sim \frac{M_\mathrm{KK} ^5 L^5 \l^2}{N_c} \sim r^{-10/3} \frac{\l^2}{N_c} \ll 1 \ ,
\ea
where we have used (\ref{conditiononmkkL}).
\subsubsection{Axion mass term and mixing with the $\etaprime$}
The combined presence of a $\theta$-term and a PQ $D8$-brane affects the equation of motion and the Bianchi identity for $\tilde F_2$ precisely as in the case of the antipodal WSS branes. Actually, in \cite{Bergman:2006xn} it has been shown that the action (\ref{SF2}) is the same for the non-antipodal configuration, with the 
obvious substitutions $\etaprime \rightarrow a$, $f_{\pi} \rightarrow f_a$
\be\label{SF2a}
S_{\tilde F_2}^{(a)}= -\frac{\chi_\mathrm{WYM}}{2}\int d^4x\left(\theta + \frac{\sqrt{2}}{f_{a}}a\right)^2\,.
\ee
From this formula it follows that, in the pure Yang-Mills case, the axion mass is given by the WV formula (\ref{mWVaxion}).

By combining together the contribution from the $N_f$ WSS $D8$-branes and that of the PQ one, we get a total action term
\be\label{SF2b}
S_{{\tilde F}_2}^{\rm{tot}}= -\frac{\chi_\mathrm{WYM}}{2}\int d^4x\left(\theta + \frac{\sqrt{2N_f}}{f_{\pi}}\etaprime + \frac{\sqrt{2}}{f_{a}}a\right)^2\,.
\ee
The total action made up summing this term, the axion kinetic term (\ref{axkin}), the effective WSS action (\ref{chiL}) without the Skyrme term and the flavor mass term (\ref{smass}) exactly reproduces the axion-dressed chiral Lagrangian (\ref{leffe}).

\subsection{Axion Lagrangian in the deconfined phase}

We now consider the Dirac-Born-Infeld action on the non-antipodal axionic PQ $D8$-brane at finite temperature, in the regime where it keep on being connected whereas the QCD quark branes are disjoint, as depicted in Figure~\ref{fig2}. Such action reads
\begin{equation}	\label{SD8T}
\tilde{S}_{D8} = -\tau_8\, \int\!d^9x\ e^{-\phi}\sqrt{\det\tilde{g}_{MN} +2\pi\alphap\,\mathcal{F}_{MN}\,}\ .
\end{equation}
The induced metric on the $D8$-brane can be derived by the ten-dimensional one~\eqref{wymT}, and reads
\begin{equation}
d\tilde{s}^2 = \Big(\bfrac{u}{R}\Big)^{\!\frac32}\bigg[-\tilde{f}(u)\,dt^2 +dx_idx^i +\left(\bfrac{1}{u'^2}+\Big(\bfrac{R}{u}\Big)^{\!3}\bfrac{1}{\tilde{f}(u)}\right)du^2\bigg] +\Big(\bfrac{R}{u}\Big)^{\!\frac32}u^2d\Omega_4^2 \ , \label{dsD8T}
\end{equation}
where $u'=\bfrac{du_{\phantom{4}}}{dx_4}$. 

\begin{figure}[ht!]
\centering
\includegraphics[scale=.6]{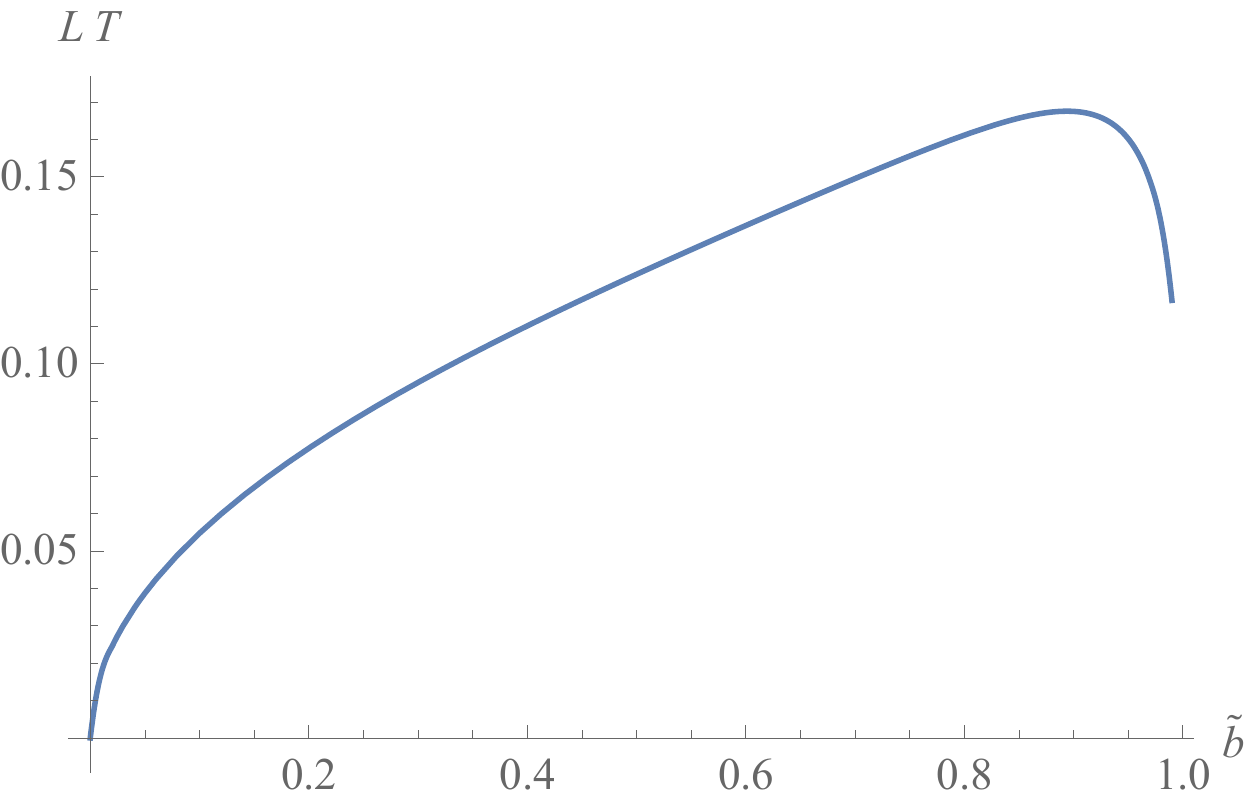}
\caption{LT as a function of $\tt{b}$}
\label{LTb}
\end{figure}

At the zeroth order we have the usual action giving the first integral (which yet differs from the one at zero temperature, because of the blackening factor moving from~$dx_4$ to~$dt$)
\begin{equation}	\label{firstintT}
u_J^4\sqrt{\tilde{f}(u_J)} = \frac{u^4\sqrt{\tilde{f}(u)}}{\sqrt{1+\big(\bfrac{R}{u}\big)^3\bfrac{{u'}^2}{\tilde{f}(u)}\,}}\ .
\end{equation}
The distance $L$ is now given by
\be
\label{distanceLfinitetemperature}
L = J_T(\tt{b}) R^{3/2} u_J^{-1/2}\,,
\ee
where $\tt{b}= u_T/u_J$ and 
\be
J_T(\tt{b}) = \frac{2}{3} \sqrt{1-\tt{b}^3\,} \int_0 ^1 dy \frac{y^{1/2}}{\sqrt{1-\tt{b}^3 y\,}\sqrt{1-\tt{b}^3 y -(1-\tt{b}^3)y^{8/3}}\,}\,.
\ee
At zero temperature, eq.s~(\ref{Lint}--\ref{elleinfunzionediuzero}), by fixing $L$ we also fix $u_J$. When we turn on the temperature $T$, 
$u_J$ becomes a non-trivial function of $T$. Indeed, recalling 
$9 u_T=16\pi^2R^3 T^2$ we have
\be
\label{LTfunctionofb}
LT = \frac{3}{4\pi}\sqrt{\tt{b}\,} J_T(\tt{b})\,.
\ee
In figure \ref{LTb} we represent $LT=LT(\tt{b})$.
Notice that the solution we are focusing on only holds for $L T \leq L T_\chi \approx 0.154$. At $T=T_{\chi}$ there is a first order phase 
transition towards a configuration where the axion is dissolved since the Peccei-Quinn $D8$-brane splits into two disconnected branches \cite{Aharony:2006da}.  
The occurrence of this transition explains the behavior of $LT(\tt{b})$ shown in figure \ref{LTb}. 
This function increases with $\tt{b}$ up to a 
maximum value $LT_m(b_m)>LT_{\chi}$ above which the connected $D8$ brane solution does not exist anymore. For $\tt{b}>\tt{b}_m$ the function decreases with 
$\tt{b}$ and this corresponds to an unstable branch.

Going back to the action~\eqref{SD8T}, at the quadratic order in~$\alphap$ we have
\begin{align}
\tilde{S}_{\mathcal{FF}} &=
	-\frac{\tau_8}{4}\left(2\pi\alphap\right)^2\, \int\!d^9x\ e^{-\phi}\sqrt{\det\tilde{g}\,}\ \tilde{g}^{MR}\tilde{g}^{NS}\mathcal{F}_{MN}\mathcal{F}_{RS} 	\nn
&=
	-C \int\!d^4x\,du\; u^{\frac52} \sqrt{\bfrac{\tilde{f}(u)}{\tilde{\gamma}(u)}\,}\ \tilde{g}^{MR}\tilde{g}^{NS}\mathcal{F}_{MN}\mathcal{F}_{RS}\ , 	\label{SD8FF}
\end{align}
where
\begin{equation}	\label{chiTc}
\tilde{\gamma}(u) = \tilde{f}(u) -\frac{u_J^8}{u^8}\,\tilde{f}(u_J)\ ,
\end{equation}
and the constant~$C$ is the same as defined in eq.~\eqref{constC}.

We then use equation~\eqref{firstintT} to express
\begin{equation}
\tilde{g}^{uu} = {u'}^2 \tilde{g}^{44} = \left(\bfrac{u}{R}\right)^{\!\frac32}\tilde{\gamma}(u)\ .
\end{equation}
Using this and the expression of the metric~\eqref{dsD8T}, we are able to expand the action~\eqref{SD8FF} as follows:
\begin{align}
\tilde{S}_{\mathcal{FF}} = -2C \int\!d^4x\,\int_{u_J}^{\infty}\!du\, \bigg\{
	\frac{R^3\,u^{-\frac12}}{\sqrt{\tilde{\gamma}(u)\tilde{f}(u)\,}}\Big[&
		-\mathcal{F}_{ti}\mathcal{F}_{ti} +\bfrac12\tilde{f}(u)\,\mathcal{F}_{ij}\mathcal{F}_{ij} \Big] \nn
	+u^{\frac52}\, \sqrt{\frac{\tilde{\gamma}(u)}{\tilde{f}(u)}\,} \Big[&
		-\mathcal{F}_{tu}\mathcal{F}_{tu} +\tilde{f}(u)\,\mathcal{F}_{iu}\mathcal{F}_{iu}\; \Big] \bigg\}\ . 	\label{SD8FFtiu}
\end{align}
Notice that the temperature breaks Lorentz invariance on the four-dimensional boundary, as known. 

We now define a suitable base of functions, such that
\begin{subequations}\begin{align}
\mathcal{F}_{tu} &= \sum_n \left(\partial_t\varphi^{(n)}(t,\vec{x})\,\tilde{\phi}_n(u) -A_t^{(n)}(t,\vec{x})\partial_u\alpha_n(u)\right)\ , \label{Ftumodes}\\
\mathcal{F}_{iu} &= \sum_n \left(\partial_i\varphi^{(n)}(t,\vec{x})\,\tilde{\phi}_n(u) -A_i^{(n)}(t,\vec{x})\partial_u\beta_n(u)\right)\ ,	\label{Fiumodes}\\
\mathcal{F}_{tj} &= \sum_n \left(\partial_tA_j^{(n)}(t,\vec{x}) -\partial_jA_t^{(n)}(t,\vec{x})\right)\alpha_n(u)\ , 				\label{Ftjmodes}\\
\mathcal{F}_{ij} &= \sum_n \left(\partial_iA_j^{(n)}(t,\vec{x}) -\partial_jA_i^{(n)}(t,\vec{x})\right)\beta_n(u)\ .					\label{Fijmodes}
\end{align}\end{subequations}
In order to recover a four-dimensional action for a massive vector field, such that it satisfies the equation of motion~$\partial_t^2A_t^{(n)}=-m_n^2A_t^{(n)}$, we have to require
\begin{subequations}\begin{align}
&
	4CR^{-\frac32}\int_{u_J}^{\infty}\!du\left(\bfrac{R}{u}\right)^{\!\frac92}\frac{u^4}{\sqrt{\tilde{\gamma}\tilde{f}\,}}\: \alpha_n\alpha_m =\delta_{nm} \ ,	\\
&\vphantom{\int_{u_J}^{\infty}}
	\Big(\bfrac{R}{u}\Big)^3\tilde{\gamma}\tilde{f}\,(\partial_u\beta_n)(\partial_u\beta_m) = m^2_n\alpha_n\alpha_m \ . 	\label{eqvecmodmass}
\end{align}\end{subequations}

The scalar sector in~(\ref{Ftumodes},~\ref{Fiumodes}) displays a zero-mode~$\tilde{\phi}_0$, that should be normal to the higher modes, which in turn can be absorbed in redefinition of the vector fields by a gauge transformation on the boundary, as it is usual fashion in the Sakai-Sugimoto model. We then require the diagonal terms in the zero-mode scalar field and the gauge fields to vanish, that is
\begin{subequations}\begin{align}
&
	4C\int_{u_J}^{\infty}\!du\ u^{\frac52}\tilde{f}^{-\frac12}\,\tilde{\gamma}^{\frac12}\; \tilde{\phi}_0\, \partial_u\alpha_n =0 \ ,	\\
&
	4C\int_{u_J}^{\infty}\!du\ u^{\frac52}\sqrt{\tilde{\gamma}\tilde{f}\,}\;  \tilde{\phi}_0\, \partial_u\beta_n =0 \ .
\end{align}\end{subequations}
These relations are satisfied by
\begin{equation}
\tilde{\phi}_0 = \bfrac{\tilde{C}_0}{2}\, u^{-\frac52}\tilde{f}^{\frac12}\,\tilde{\gamma}^{-\frac12}\ , 	\label{phi0}
\end{equation}
(with $\tilde{C}_0$ a so far undetermined constant) if we take
\begin{equation}
\partial_u\alpha_n \equiv \tilde{f}\,\partial_u\beta_n \ .
\end{equation}
With this last identity the equation~\eqref{eqvecmodmass} can be rewritten into a differential equation for the modes~$\alpha_n$'s by themselves:
\begin{equation}
\partial_u\Big[\Big(\bfrac{R}{u}\Big)^{\!3}\tilde{f}^{-1}\tilde{\gamma}\, \partial_u\alpha_n\Big] +m_n^2\alpha_n =0\ .
\end{equation}

Now we have to normalize the zero-mode~$\tilde{\phi}_0$ in order to have a canonically normalized kinetic term~$\frac12\big(\partial_t\varphi^{(0)}\big)^2$ in the four-dimensional 
Lagrangian. This is given by
\begin{equation}
4C\int_{u_J}^{\infty}\!du\ u^{\frac52}\tilde{f}^{-\frac12}\,\tilde{\gamma}^{\frac12}\;  \tilde{\phi}_0^2 \equiv
	C\tilde{C}_0^2\int_{u_J}^{\infty}\!du\ u^{-\frac52}\tilde{f}^{\frac12}\,\tilde{\gamma}^{-\frac12} =1 \ , 	\label{normalphi0}
\end{equation}
which thus fixes the constant~$\tilde{C}_0$.

Using all these orthonormality relations, we can write down the part of the action~\eqref{SD8FFtiu} that concerns the scalar mode~$\varphi^{(0)}$, that is the axion,
\begin{equation}
\int\!d^4x \left[ \bfrac12\partial_t\varphi^{(0)}\partial_t\varphi^{(0)} -\bigg(1-u_T^3\;C\tilde{C}_0^2\int_{u_J}^{\infty}\!du\:u^{-\frac{11}2}\,\tilde{f}^{\frac12}\,\tilde{\gamma}^{-\frac12}\bigg)\bfrac12\partial_i\varphi^{(0)}\partial_i\varphi^{(0)} \right]\ ,
\end{equation}
where we have used the explicit expression of the thermal form-factor~$\tilde{f}$ in \eqref{wymT} in order to highlight the dependence on the temperature through~$u_T$, which deforms the dispersion relation of the axion from the relativistic one, {\it i.e.}:
\begin{equation}	\label{disrelax}
\omega_k^2 = \left(1-u_T^3\;C\tilde{C}_0^2\int_{u_J}^{\infty}\!du\:u^{-\frac{11}2}\,\tilde{f}^{\frac12}\,\tilde{\gamma}^{-\frac12}\right) k^2\ .
\end{equation}
For $b \ll1 $, using (\ref{LTfunctionofb}), we obtain
\begin{align}	\label{disrelaxnum}
\omega_k^2 &\sim
	\left(1-\frac{16 \Gamma \pr{ \tfrac{9}{16}} \Gamma \pr{ \tfrac{11}{16} } }{\Gamma \pr{ \tfrac{1}{16} } \Gamma \pr{ \tfrac{3}{16}}}\; b^3\right) k^2 \nn
&\simeq
	\pq{ 1-16383.2\; L^6 T^6 + \cdots} k^2 \,.
\end{align}
Analogously, we can wonder what is the effect of the temperature on the axion decay constant~$f_a$. To this purpose, we apply the same reasoning of equations~\eqref{holonomyV} and~\eqref{V1storder} to the current finite-temperature model, and we obtain
\begin{equation}
V \sim 	1 +2i\int_{u_J}^{\infty}\!du\, \varphi^{(0)}\tilde{\phi}_0 = 
					1 +i\varphi^{(0)}\; \tilde{C}_0\int_{u_J}^{\infty}\!du\ u^{-\frac52}\tilde{f}^{\frac12}\,\tilde{\gamma}^{-\frac12} =
					1 +\frac{i}{C\tilde{C}_0}\,\varphi^{(0)}\ ,
\end{equation}
where in the last step we have used the normality relation~\eqref{normalphi0}. Then, having in mind the definition~\eqref{holonomyV}, we straightforwardly read the temperature-dependent decay constant
\begin{equation}\label{faTcc0}
f_a^2(T) = 2C^2\tilde{C}_0^2 \equiv
	2C\left(\int_{u_J}^{\infty}\!du\ u^{-\frac52}\tilde{f}^{\frac12}\,\tilde{\gamma}^{-\frac12}\right)^{\!-1}\ .
\end{equation}
In terms of the coordinate $z$, defined by $u^3 = u_J^3 + u_J z^2$, this result reads
\be
\label{faTcc0coordinatez}
f_a^2(T) = \frac{3C}{u_J} \left(\int_{0}^{\infty}\!dz\ z u^{-\frac92}\tilde{f}^{\frac12}\,\tilde{\gamma}^{-\frac12}\right)^{\!-1}\ .
\ee
This result agrees with the one that was found in~\cite{Aharony:2007uu} by another argument.

\begin{figure}[ht!]
\centering
\includegraphics[scale=.6]{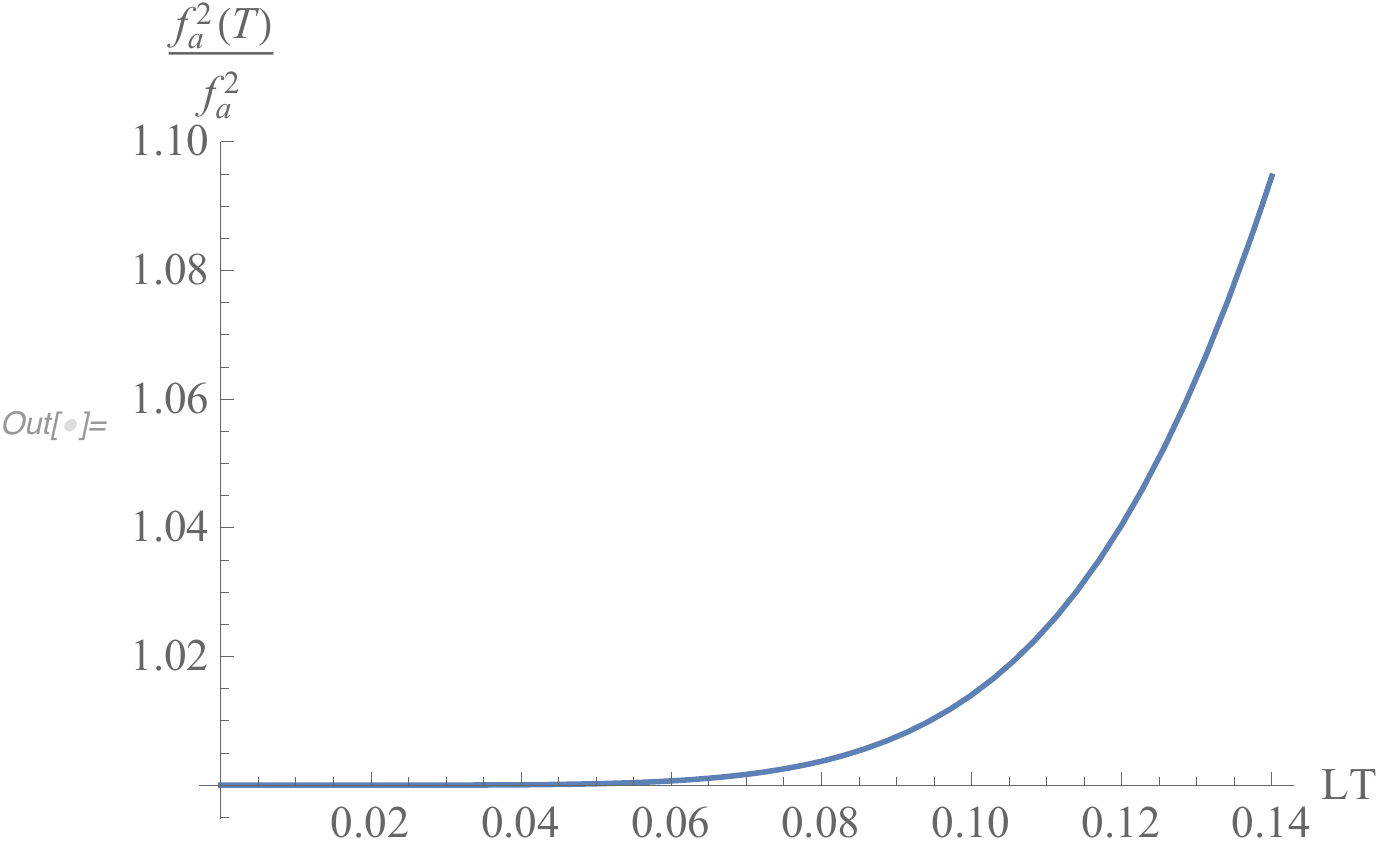}
\caption{$f_a^2(T)$ as a function of $LT$}
\label{fa2T}
\end{figure}

In terms of field theory quantities (and $L$) the axion decay constant at finite temperature reads
\be
f_a ^2 (T)= 6C\,\frac{u_J^\frac32}{I_T(\tilde{b})} =\frac{N_c \l}{16 \pi^3 } \frac{J_T^3(\tt{b})}{I_T(\tt{b})}  \frac{1}{M_\mathrm{KK} L^3}
\ ,
\ee
where
\be
I_T(\tt{b})=\int_{0}^{1}dy \frac{y^{-\frac12}\sqrt{1-\tt{b}^3y\,}}{\sqrt{1-\tt{b}^3y -y^{\frac83}(1-\tt{b}^3)\,}}\ .
\ee
A plot of $f_a^2(T)$ as a function of $LT$ is given in figure \ref{fa2T}.
For $LT\ll 1$ we get
\be
f_a^2(T)\approx f_a^2\left[1+14247\, L^6 T^6+\cdots \right]\,,
\ee
where the zero temperature axion coupling $f_a$ is given in (\ref{fa0bsmall}).
The leading order term coincides with the one we found at zero temperature in the limit $u_J\gg u_0$, see eq. (\ref{fa0bsmall}).
\section{Non-local Nambu-Jona Lasinio model}\label{appendix:njl}
From the field theory point of view, the Witten-Sakai-Sugimoto model is a five-dimensional gauge theory with two Weyl fermions 
of opposite chiralities placed at two antipodal points along the fifth, compactified $x_4$-dimension. In this sense, the fermions are 
co-dimension one defects. In the main text we extended the model in order to include the Peccei-Quinn axion. This has been 
achieved by the introduction of an additional pair of $D8$-branes, which amounts to adding two additional co-dimension one defects 
in the five-dimensional field theory.  In order to disentangle the chiral symmetry breaking mechanism (whose associated Nambu-Goldstone boson 
we identified with the axion) from confinement, the additional Weyl fermions had to be taken in a non-antipodal configuration and in the 
regime $L \ll \pi R_4$, where $L$ is the distance between fermions along the $x_4$ direction and $R_4$ is the compactification radius. 

In the limit in which $R_4 \rightarrow \infty$ the gauge theory with two non-antipodal fermions reduces 
to the so-called non-local Nambu-Jona Lasinio model studied in 
\cite{Antonyan:2006vw}.
It is not straightforward to extrapolate the considerations in \cite{Antonyan:2006vw} to the WSS confining case.
If the separation $L$ of the $D8/\overline{D8}$-branes is small compared to the length of the semi-circle, then the analysis is the same as in \cite{Antonyan:2006vw} to a very good accuracy.
In \cite{Dhar:2009gf} there is a first field theory analysis of the effects of the compact circle, but still in the $L \ll \pi R_4$ case.
The idea is that now the theory has a four dimensional phase in the far IR, where the non-local model should reduce to the local one.
As such, chiral symmetry breaking should be present only above a certain value of the coupling.
A concrete realization of this idea is far from established.

In this appendix we shall follow the analysis in \cite{Antonyan:2006vw} in order to extract information about the 
relative strengths of the interactions among the fermions. We shall consider the gauge theory in the linear approximation, 
hence treating the gauge field as an Abelian field and integrating it out. Since the gauge field is massless, its integration
yields a Lagrangian with non-local quartic fermion interactions. We shall first consider the case with only one 
defect, then with two defects and finally with four defects, as in the WSS model generalized by the inclusion of the axion. 
The analysis is meant to be valid only in the 
case in which the distance between two fermions is much smaller than the radius of compactification of the fifth direction. 
This forces us to take in a non-antipodal configuration also the QCD quarks.
Moreover, due to the non-locality of the 
model, we have to limit ourselves to the massless fermions case, as will be explained at the end of this appendix.

\subsubsection*{One defect case}

Let us consider the gauge theory with a set of $N_f$ left Weyl fermions $q_{L} ^i$ placed 
at $x_4 =0$, where $i$ is a flavor index. The system is invariant under global $\mathrm{U}(N_f)_L$ transformations which act as
\be
q_L ^i \rightarrow (U_L)^{i} _j q_L ^j \ , \q  \q \q U_L \in \mathrm{U}(N_f)_L \ .
\ee
In the following we will omit the flavor indices. Capital latin letters will label the whole set of coordinates whereas greek 
letters will label all the coordinates except $x_4$. The classical action is 
\be
\label{njlmodeloneflavor}
S = \int d^5x \pq{- \frac{1}{4 g_5 ^2} F_{MN} ^2 + \d (x_4) q_L ^\dagger \bar{\s} ^\m \pr{ i \p_\m + A_\m} q_L } \ .
\ee
Here we use the standard notation $\s^\m = (\mathrm{1},\s^i)$,  $\bar{\s}^\m = (\mathrm{1},-\s^i)$,where $\s^i$ are the Pauli matrices.
The gauge part of the Lagrangian is the Maxwell Lagrangian with a matter current
\be
j^M(x,x_4) = \d^M _\m \d (x_4) q_L ^\dagger (x) \bar{\s} ^\m q_L (x)
\ee
which transforms in the adjoint of the gauge group and it is a singlet under the global group $\mathrm{U}(N_f)_L$.
Let us integrate out the gauge field. In the Feynman gauge the propagator for the Abelian gauge field reads
\be
\tt{\D} _{MN} (p) = g_5 ^2 \tt{G} (p) \eta_{MN}= - \frac{i g_5 ^2}{p^2} \eta_{MN} \ .
\ee
In coordinate space the propagator $G(x)$ reads
\ba
G(x,x_4) = \frac{1}{8 \pi^2} \frac{1}{| \eta_{\m \n} x^\m x^\n + x_4 ^2|^{3/2}} \ .
\ea 
We can use the propagator in order to express the gauge field in an integral form:
\ba
A_M(x,x_4) =  -g_5 ^2 \int d^5 y G (x-y,x_4-y_4) \eta_{\m M} \d (y_4) q_L ^\dagger (y)  \bar{\s} ^\m q_L(y)  \ . \q 
\ea
Inserting this result in (\ref{njlmodeloneflavor}) the interaction term becomes\footnote{Here we use the Fierz identity
\be
\pr{ \psi ^\dagger _{1L} \bar{\s} ^\m \psi _{2L} } \pr{ \psi ^\dagger _{3L} \bar{\s} _\m \psi _{4L} } = 
\pr{ \psi ^\dagger _{1L} \bar{\s} ^\m \psi _{4L} } \pr{ \psi ^\dagger _{3L} \bar{\s} _\m \psi _{2L} } \ .
\ee}
\ba
\label{njlinteractiontermoneflavor}
S_\text{int} = - g_5 ^2 \int d^4 x d^4 y G(x-y,0) \pq{ q_L ^\dagger (x)  \bar{\s} _\m q_L(y) } \pq{ q_L ^\dagger (y)  \bar{\s} ^\m q_L(x) } \ . \q 
\ea
The terms in square bracket in (\ref{njlinteractiontermoneflavor}) are singlet under the gauge group and 
transform in the adjoint of the global flavor group.
The interaction term $S_\text{int}$ gives a correction to the free field propagator.

\subsubsection*{Two defects}

Let us now consider the case in which we have $N_f$ left Weyl fermions placed at $x_4=L/2$ and 
$N_f$ right Weyl fermions at $x_4=-L/2$. In addition to the global symmetry $\mathrm{U}(N_f)_L$ there is a symmetry under
the global transformations $\mathrm{U}(N_f)_R$  which act as
\be
q_R ^i \rightarrow (U_R)^{i} _j q_R ^j \ , \q  \q \q U_R \in \mathrm{U}(N_f)_R \ .
\ee
The classical action is then
\ba
\label{njlmodeltwoflavor}
S &=& \int d^5x \bigg[- \frac{1}{4 g_5 ^2} F_{MN} ^2 + \d \pr{x_4- L/2} q_L ^\dagger \bar{\s} ^\m \pr{ i \p_\m + A_\m} q_L + \nb \\
&+& \d (x_4+L/2) q_R ^\dagger \s ^\m \pr{i \p_\m + A_\m} q_R  \bigg] \ .
\ea
In this case the matter current is
\be
j^M(x,x_4) = \d^M _\m \d (x_4-L/2) q_L ^\dagger (x) \bar{\s} ^\m q_L (x) + \d^M _\m \d (x_4+L/2) 
q_R ^\dagger (x) \s ^\m q_R (x) \ ,
\ee
hence
\ba
A_M(x,x_4)
&=&- g_5 ^2 \int d^5 y G (x-y,x_4-y_4) \eta_{\m M} 
\Big(\d (y_4-L/2) q_L ^\dagger (y)  \bar{\s} ^\m q_L(y)+ \nb \\
&+& \d (y_4+L/2) q_R ^\dagger (y)  \s ^\m q_R(y) \Big) \ .
\ea
Inserting this expression in the action (\ref{njlmodeltwoflavor}) we are left with three interaction terms: two of them give 
corrections to the propagator of the left and right fermions, akin to the term derived in the previous subsection; the other 
one describes the interaction between left and right fermions and reads\footnote{In order to obtain this term we use the Fierz identity
\be
\pr{ \psi_{1L} ^\dagger \bar{\s}^\m \psi_{2L} } \pr{ \psi_{3R} ^\dagger \s_\m \psi_{4R} } = 2 \pr{\psi_{1L}^\dagger \psi_{4R}}
\pr{\psi^\dagger_{3L} \psi_{2R}} \ .
\ee}
\ba
\label{njlinteractiontermtwoflavors}
S_\text{int} =-4 g_5 ^2 \int d^4 x d^4 y G (x-y,L) \pq{  q_L ^\dagger (x) \cdot q_R(y) } \pq{ q_R ^\dagger (y) \cdot q_L(x)}   \ .
\ea
Each term in square bracket is a singlet under the gauge group and transforms 
in the adjoint under $\mathrm{U}(N_f)_L \times \mathrm{U}(N_f)_R$.

\subsubsection*{Four defects}

Let us consider the case in which we have two ``quark defects'' separated by a distance $L'$ as in the previous subsection and 
two ``axion defects'' separated by a distance $L\ll L'$.
In particular we will consider a single left Weyl fermion $\psi_L (x)$
placed at $x_4 = -L/2$ and a single right Weyl fermion $\psi_R (x)$ placed at $x_4 = L/2$.
In addition to the $\mathrm{U}(N_f)_L \times \mathrm{U}(N_f)_R$ there is also a symmetry under the $\mathrm{U}(1)_L \times \mathrm{U}(1)_R$ transformations which act 
on  the axion quarks. The action reads
\ba
\label{njlmodelfourflavors}
S &=& \int d^5x \bigg[- \frac{1}{4 g_5 ^2} F_{MN} ^2 + \d \pr{x_4- L'/2} q_L ^\dagger \bar{\s} ^\m \pr{ i \p_\m + A_\m} q_L  +\nb \\
&+& \d \pr{x_4 + L/2} \psi_L ^\dagger \bar{\s} ^\m \pr{i \p_\m + A_\m} \psi_L + 
 \d (x_4+L'/2) q_R ^\dagger \s ^\m \pr{i \p_\m + A_\m} q_R   \nb \\
&+& \d \pr{x_4- L/2} \psi_R ^\dagger \s ^\m \pr{i \p_\m + A_\m} \psi_R \bigg] \ .
\ea
The matter current in this case is
\ba
j^M(x,x_4) &=& \d^M _\m \d (x_4-L'/2) q_L ^\dagger (x) \bar{\s} ^\m q_L (x) + \d^M _\m \d (x_4+L'/2) 
q_R ^\dagger (x) \s ^\m q_R (x) + \nb \\
&+&\d^M _\m \d (x_4+L/2) \psi_L ^\dagger (x) \bar{\s} ^\m \psi_L (x) + \d^M _\m \d (x_4-L/2) 
\psi_R ^\dagger (x) \s ^\m \psi_R (x)  \ ,  \nb
\ea
hence the gauge field can be written as
\ba
A_M(x,x_4)
&=&- g_5 ^2 \int d^5 y G (x-y,x_4-y_4) \eta_{M \m} 
\Big(\d (y_4-L'/2) q_L ^\dagger (y)  \bar{\s} ^\m q_L(y)+ \nb \\
&+& \d (y_4+L/2) \psi_L ^\dagger (y)  \bar{\s} ^\m \psi_L(y) + \d (y_4+L'/2) q_R ^\dagger (y)  \s ^\m q_R(y) + \nb \\
&+& \d (y_4-L/2) \psi_R ^\dagger (y)  \s ^\m \psi_R(y) \Big) \ .
\ea
Plugging this expression in (\ref{njlmodelfourflavors}) we  obtain four kinds of terms:
\begin{itemize}
\item quartic terms which give corrections to the fermion propagators;
\item quartic interactions involving both left and right quarks:
\be
S_\text{int} ^{q}=-4  g_5 ^2 \int d^4 x d^4 y G (x-y,L') \pq{  q_L ^\dagger (x) \cdot q_R(y) } \pq{ q_R ^\dagger (y) \cdot q_L(x)}   \ ;
\ee
\item quartic interactions involving both left and right axion quarks:
\be
S_\text{int} ^{ax}=-4  g_5 ^2 \int d^4 x d^4 y G (x-y,L) \pq{  \psi_L ^\dagger (x) \cdot \psi_R(y) } 
\pq{ \psi_R ^\dagger (y) \cdot \psi_L(x)}  \ ;
\ee
\item quartic interactions which involve two quarks and two axion quarks:
\ba
S_\text{int} ^{q,ax}&=&- 4  g_5 ^2 \int d^4 x d^4 y G(x-y,L'/2-L/2) \Bigg\{ \pq{  q_L ^\dagger (x) \cdot \psi_R(y) } \times \nb \\
&\times&  \pq{ \psi_R ^\dagger (y) \cdot q_L(x)}
+ \pq{ \psi_L ^\dagger (y) \cdot q_R(x)} \pq{ q_R ^\dagger (x) \cdot \psi_L(y)} \Bigg\} + \nb \\
&-& 2 g_5 ^2 \int d^4 x d^4 y G(x-y,L'/2 + L/2) \Bigg\{ \pq{ q_L ^\dagger (x)  \bar{\s} ^\m \psi_L(y) } \times \nb \\
&\times&  \pq{ \psi_L ^\dagger (y)  \bar{\s} _\m q_L(x) }
+ \pq{ q_R ^\dagger (x)  \s ^\m \psi_R(y) } \pq{ \psi_R ^\dagger (y)  \s _\m q_R(x) } \Bigg\}  \q \q \q 
\ea
\end{itemize}
We expect the relative strengths between the different interactions to depend on the distance between the fields along the fifth direction. 
Hence, in order to estimate them let us consider
\be
G(x-y,L) \sim \frac{1}{L^3} \ .
\ee
Recalling that $4\pi g_5^2 = M_\mathrm{KK} g_\mathrm{YM}^2$ and the result (\ref{fares}) for the axion decay constant, we have
\be
S_\text{int} ^{ax} \propto \frac{g_5 ^2}{L^3}  \sim \frac{M_\mathrm{KK} ^2 }{N_c ^2} f_a ^2 \ .
\ee
Analogously,
\be
S_\text{int} ^{q} \propto \frac{g_5 ^2}{L'^3}  \sim \frac{M_\mathrm{KK} ^2}{N_c ^2} f_q ^2 \ ,
\ee
where 
\be
f_q ^2 = \frac{1}{16 \pi^3} \frac{J^3(b)}{I(b)}  \frac{\l}{M_\mathrm{KK} L'^3} N_c \ .
\ee
In the regime $L \ll L'$,
\be
S_\text{int} ^{q,ax} \propto \frac{g_5 ^2}{L^3}  \sim \frac{M_\mathrm{KK} ^2}{N_c ^2} f_q ^2 \ .
\ee
Thus, we expect the interaction involving two quarks and two axion quarks to be as relevant as the interactions between the quarks.

Finally, let us stress the fact that in this appendix we have been discussing the non-local NJL model assuming the quarks to be massless. Due to the non-locality of the model, 
we cannot add a Dirac mass term to the Lagrangian. The best we can do is to consider a Wilson line such as
\be
S_\text{mass} =- m \int d^4 x d^4 y q_R ^\dagger (x) \mathcal{P} \exp \pr{ i \int _{-L/2} ^{L/2} d x_4 A_4 } q_L (y) \ .
\ee
Unfortunately, such a term makes the gauge field equations non-linear, thus preventing us from proceeding with the analysis.


\end{document}